\documentclass[aps,twocolumn,showpacs]{revtex4}
\usepackage{graphicx}
\usepackage[all]{xy}
\usepackage{amsmath}
\usepackage{amssymb}
\usepackage[latin1]{inputenc}
\newcommand{\be}{\begin{equation}}
\newcommand{\ee}{\end{equation}}
\newcommand{\ben}{\begin{eqnarray}}
\newcommand{\een}{\end{eqnarray}}
\newcommand{\bes}{\begin{subequations}}
\newcommand{\ees}{\end{subequations}}

\newcommand{\bb}{\bibitem}

\begin{document} 
\title{New Lump-like Structures in Scalar-field Models}
\author{A.T. Avelar$^a$,  D. Bazeia$^b$, L. Losano$^b$, and R. Menezes$^b$}
\affiliation{{\small $^a$Instituto de F\'\i sica, Universidade Federal de Goi\'as, 74.001-970 Goi\^ania, Goi\'as, Brazil}
\\{\small $^b$Departamento de F\'{\i}sica, Universidade Federal da Para\'\i ba, 58051-970 Jo\~ao Pessoa, Para\'\i ba, Brazil}}

\begin{abstract}
In this work we investigate lump-like solutions in models described by a single real scalar field. We start considering non-topological solutions with the usual lump-like form, and then we study other models, where the bell-shape profile may have varying amplitude and width, or develop a flat plateau at its
top, or even induce a lump on top of another lump. We suggest possible applications where these exotic solutions might be used in several distinct branches of physics.
\end{abstract}

\maketitle

\section{Introduction}
\label{s1}

Defect structures are of great interest in physics in general. They are classical configurations of topological or non-topological profile, and have been the subject of many investigations both in high energy \cite{R,W,VS,MS} and in several other areas of physics \cite{KdV,M,dav,mur,wal,agra}. In relativistic field theory in general, the topological defects are linearly stable while the non-topological ones are unstable.

Topological or kink-like defects are of interest in many different contexts, in particular in soft condensed matter, where they may appear as interfaces between distinct regions to contribute to pattern formation \cite{mur,wal}, in optics communications, as dark solitons in fibers \cite{agra,kivshar} and in cosmology, to induce the formation of structure in the early Universe \cite{VS,MS}, and more recently to model dark energy \cite{BS,MM}. They are also used in many different contexts in the brane world scenario involving a single extra dimension \cite{ADD,RS,GW,DF}.

Non-topological or lump-like structures also appear in several contexts in physics. The non-topological profile is of the bell-shape form, as it is, for instance, the well-known soliton of the KdV equation \cite{KdV}. They are of current interest in soft condensed matter \cite{mur,wal,dav} to describe, for instance, charge transport in diatomic chains \cite{P1,Xu,Xu1,bnt,bllm}, in optics communication, to describe bright solitons in fibers \cite{agra,HW}, and in high energy physics in diverse contexts as, for instance, seeds for the formation of structures \cite{FG,MP,MB,CL,kh}, q-balls \cite{C,DK,K,Ma}, tachyon brane \cite{S,zwi1,zwi2}, and galactic dark matter properties \cite{m1,m2}. They can also be used to describe brane world scenarios with a single extra dimension \cite{Sto,MG}, and we hope that the present investigations will provide a new route on the subject.

In applications where the non-topological structures may play a role, they usually have standard bell-shape profile. However, the bell-shape solution may also appear with exotic profile, and this has inspired us to study non-topological structures where the modified bell-shape profile appears as intrinsic feature of the system. The study has led us with interesting results, which we report in two distinct sections. In Sec.~\ref{sec:gen}, we review the standard case. There we take advantage of the first-order formalism and we modify the usual procedure in order to offer a new method to study non-topological solutions. The existence of a first-order framework for lumps has importantly simplified the investigation on exotic solutions, and in Sec.~\ref{sec:spe} we further illustrate the procedure with three distinct generalizations, where we show how to get exotic modifications of the bell-shape profile. We end the work in Sec.~IV, with some comments and conclusions.

\section{The framework}
\label{sec:gen}

Let us consider a single real scalar field in $(1,1)$ space-time dimensions. The Lagrange density is given by
\be\label{sm}
{\cal L}=\frac12\partial_\mu\phi\,\partial^\mu\phi-V(\phi)
\ee
where $V(\phi)$ is the potential, $x^\mu=(t,x)$ and $x_\mu=(t,-x),$ and we work with dimensionless fields and coordinates. We suppose that the potential engenders a set of critical points, $\{\bar\phi_1,...,\bar\phi_n \},$ such that $V^{\prime}(\bar\phi_i)=0$ and $V(\bar\phi_i)=0$ for $i=1,2,...,n$.
The equation of motion which follows from the above model is
\be
\frac{\partial^2\phi}{\partial t^2}-\frac{\partial^2\phi}{\partial x^2}+\frac{dV}{d\phi}=0
\ee
We suppose that $\phi=\phi(x)$ is static field; thus, the equation of motion leads to 
\be\label{em1}
\frac{d^2\phi}{dx^2}=V^{\prime}(\phi) 
\ee
where the prime stands for derivative with respect to the argument. We integrate the equation of motion \eqref{em1} to get
\be\label{em1phi}
\frac{d\phi}{dx}=\pm\sqrt{2V+c}
\ee
where $c$ is a real constant.

We are searching for defect structures, and the energy density of the static solutions $\phi=\phi(x)$ is given by
\be
\epsilon=\frac12\left(\frac{d\phi}{dx}\right)^2+V(\phi)
\ee
It is the addition of two portions, known as the gradient and potential contributions, respectively. Thus, the static solutions have to obey the boundary conditions
\be
\lim_{x\to\pm\infty}\frac{d\phi}{dx}\to0
\ee
in order to ensure finiteness of the gradient portion of the energy.

The presence of the set of critical points allows to distinguish two kinds of defect structures: there may be topological or kink-like structures, which in general connect two distinct but adjacent critical points, $\bar\phi_i$ and $\bar\phi_{i+1}$ with the boundary conditions
\be
\lim_{x\to\pm\infty}\phi(x)\to\bar\phi_i\;\;{\&}\;\;\lim_{x\to\mp\infty}\phi(x)\to\bar\phi_{i+1}
\ee
and non-topological or lump-like structures, which in general requires a single critical point, with the boundary conditions
\be
\lim_{x\to\pm\infty}\phi(x)\to\bar\phi_i
\ee
These boundary conditions imply that $c$ in \eqref{em1phi} should vanish, as a necessary condition for the presence of finite energy solutions,
leading to 
\be\label{em1phi0}
\frac{d\phi}{dx}=\pm\sqrt{2V}
\ee

There are two first-order equations in \eqref{em1phi0}, and they should be considered with care. The general issue is that the topological or kink-like solutions are monotonic functions of $x$, so the derivative does not change sign and in \eqref{em1phi0} each sign identifies one equation. These two equations lead to two distinct solutions, which we name kink and anti-kink, respectively. We recall that naming kink or anti-kink is a matter of convention, and we use kink and anti-kink for positive and negative sign, respectively. However, the non-topological or lump-like solutions are trickier, because they are not monotonic. In fact, their first derivatives change sign at some arbitrary point which we name $x_0,$ the center of the solution. Due to the translational invariance of the theory we take $x_0=0,$ for simplicity. Thus, the derivative of the lump-like solutions change sign at $x=0,$ and so we have to understand the above equations \eqref{em1phi0} as
\be
\frac{d\phi}{dx}=\sqrt{2V}\;{\rm for}\; x>0\;\;{\&}\;\;\frac{d\phi}{dx}=-\sqrt{2V}\;{\rm for}\; x<0 
\ee 
and/or  
\be
\frac{d\phi}{dx}=-\sqrt{2V}\;{\rm for}\; x>0\;\;{\&}\;\;\frac{d\phi}{dx}=\sqrt{2V}\;{\rm for}\; x<0 
\ee 
The presence of the and/or connection between the above pairs of equations can be understood with the help of the reflection symmetry: the two pairs of equations are required when the model engenders reflection symmetry; otherwise, we will only need a single pair of equations -- see below for further comments on this issue.

The fact that the kink-like solutions are monotonic functions induces an important feature to the topological structures: in the study of stability, we know that the derivative of the defect structure represents the zero mode of the related quantum-mechanical problem; thus, because the topological structures are monotonic, the zero-modes are necessarily node-less, and so the quantum-mechanical problem admits no negative bound states, making the topological structures stable. The case of non-topological or lump-like structures is different: the quantum mechanical problem engenders zero-mode with a node at the center of the structure, showing that there is at least one bound state with negative energy, which induces instability.

The topological or kink-like structures are protected against instability, and this may be translated into a very interesting feature: we suppose that the potential can be written in terms of another function, $W=W(\phi),$ in a way such that
\be\label{susyp}
V(\phi)=\frac12\left(\frac{dW}{d\phi}\right)^2=\frac12 W^2_\phi
\ee
and the two first-order equations \eqref{em1phi0} can be written as
\be\label{foT}
\frac{d\phi}{dx}= W_\phi\;\;\;\;{\&}\;\;\;\;\frac{d\phi}{dx}=-W_\phi
\ee
for the kink and anti-kink, respectively. In this case, the energy can be written as
\be
E_{\rm kink}=|W(\phi(\infty))-W(\phi(-\infty))|
\ee

The form of the potential \eqref{susyp} is very interesting. Besides the simplicity of the energy calculation which we have just shown, it also leads to two other important results. The first one is that if we couple the scalar field with fermions, we can choose the Yukawa coupling as $W_{\phi\phi}.$ This induces degeneracy between boson and fermion masses, a necessary condition to make the model super-symmetric \cite{susy} -- see \cite{su} for an explicitly solvable model. The behavior of $W_{\phi\phi}(\phi(x))$, calculated at the static topological solution $\phi(x)$, is then directly connected with the presence of super-symmetry in the topological sector of the model. The second result is related to stability. In the study of classical or linear stability one usually obtains the Schroedinger-like Hamiltonian
\be\label{ha}
H=-\frac{d^2}{dx^2}+U(x);\;\;\;\;\;U(x)=\frac{d^2V}{d\phi^2}
\ee
In the topological sector, $U(x)$ is the second derivative of the potential, calculated at $\phi=\phi(x),$ that is, at the classical static solution. However, for the potential \eqref{susyp} we get
\be\label{pq}
U(x)=W^2_{\phi\phi}+W_\phi W_{\phi\phi\phi}
\ee
Thus, linear stability of the topological solution is also connected with the behavior of $W_{\phi\phi}$ at the topological sector where $\phi=\phi(x)$.
This is better seen as follows: for solutions which solve the first-order equations \eqref{foT}, we can use the Hamiltonian \eqref{ha} to write the expression $H=S^\dag S,$ where the first-order operators $S^\dag$ and $S$ are given by $S^\dag=\pm d/dx+W_{\phi\phi},$ and $S=\mp d/dx+W_{\phi\phi},$ for kink (upper sign) and anti-kink (lower sign), according to the above conventions. Thus, $H$ is non-negative and this proves that the system supports no bound state with negative eigenvalue.

This is not possible for non-topological or lump-like solutions in general. However, we can change this scenario, changing the way we investigate lump-like solutions. To do this we first recognize that the lump-like structure is monotonic for $x$ positive, and for $x$ negative, separately. Thus, for $x$ positive or for $x$ negative the non-topological profile is similar to the topological one, and so we can use symmetry arguments to introduce $W=W(\phi)$ for non-topological solutions, with the understanding that the equations are now changed to
\be\label{EM11}
\frac{d\phi}{dx}=W_\phi\;{\rm for}\; x>0\;\;{\&}\;\;\frac{d\phi}{dx}=-W_\phi\;{\rm for}\; x<0
\ee 
and/or
\be\label{EM12}
\frac{d\phi}{dx}=-W_\phi\;{\rm for}\; x>0\;\;{\&}\;\;\frac{d\phi}{dx}=W_\phi\;{\rm for}\; x<0
\ee
In this case, the energy associated with the lump-like solutions has the form
\ben\label{EL}
E_{\rm l}&\!\!=\!\!&|W(\phi(\infty))\!-\!W(\phi(0))|\!+\!|W(\phi(0))\!-\!W(\phi(-\infty))|\nonumber
\\
&=&2|W(\phi(\infty))-W(\phi(0))|\nonumber
\\
&=&2|W(\phi(0))-W(\phi(-\infty))|
\een
We call this the first-order framework for non-topological or lump-like structures. It has two important advantages over the usual calculation: the first is that it allows to obtain the energy in a simple and direct manner, which does not depend of the explicit form of the solutions, as it is required in the usual case; the second advantage is that one has to deal with first-order equations, instead of the second order equations of motion.

However, the non-topological solutions are in general unstable, so we have to investigate how this appears in the above framework. The issue is directly related to $W_{\phi\phi}$, which appears in the investigation of stability. In general, $W_{\phi\phi}$ is divergent at the center of the lump-like structure. Thus, if one tries to add fermion to construct super-symmetric extension of the model, the divergence of $W_{\phi\phi}$ makes the fermion mass ill-defined at the center of the lump-like structure. The argument which shows the divergence of $W_{\phi\phi}$ at the center of the solution can be constructed as follows: from the first-order formalism, the zero mode $\eta_0(x)$ has to obey
\be
\frac{d}{dx}\ln(\eta_0)=W_{\phi\phi}
\ee
If the zero mode has a node at the center of the solution, it has to change sign at that point, and so the logarithm cannot be continuous there, necessarily inducing a divergence in $W_{\phi\phi}$ at the center of the non topological structure.   

With this first-order framework at hand, let us now illustrate how it works, explicitly investigating some known examples. Firstly, we consider the $\phi^3$ model, which is described by the potential
\be\label{phi3}
V(\phi)=2\phi^2(1-\phi)
\ee
which is plotted in Fig.~1. The local maximum is at $\phi_{max}=2/3$. This potential breaks the reflection symmetry $\phi\to-\phi.$ However, we can introduce another model, with the potential $V(\phi)=2\phi^2(1+\phi),$ which can be seen as a partner model, connected by the reflection symmetry.

\begin{figure}[ht!]
\includegraphics[{width=5cm}]{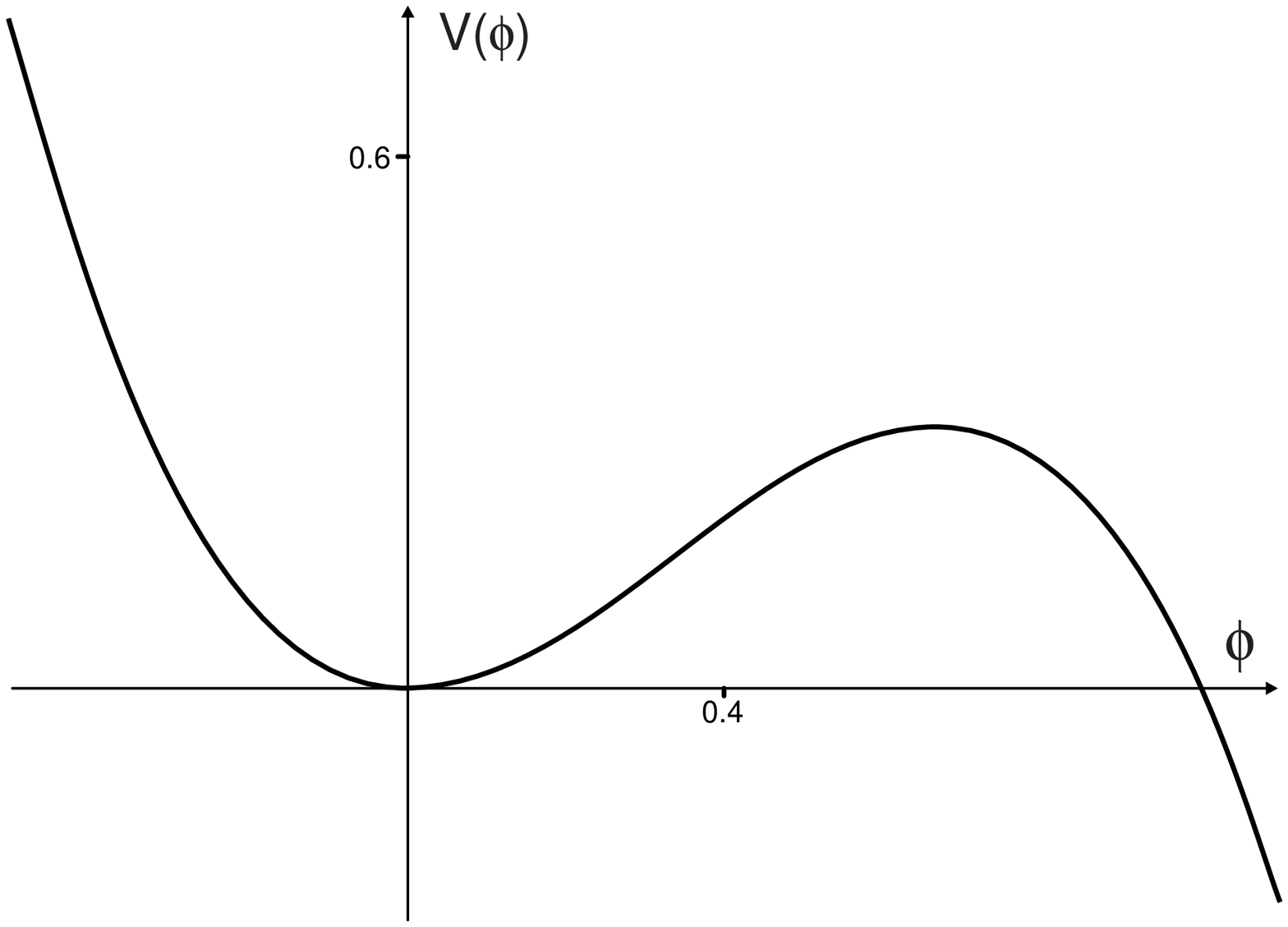}\vspace{.4cm}
\includegraphics[{width=5cm}]{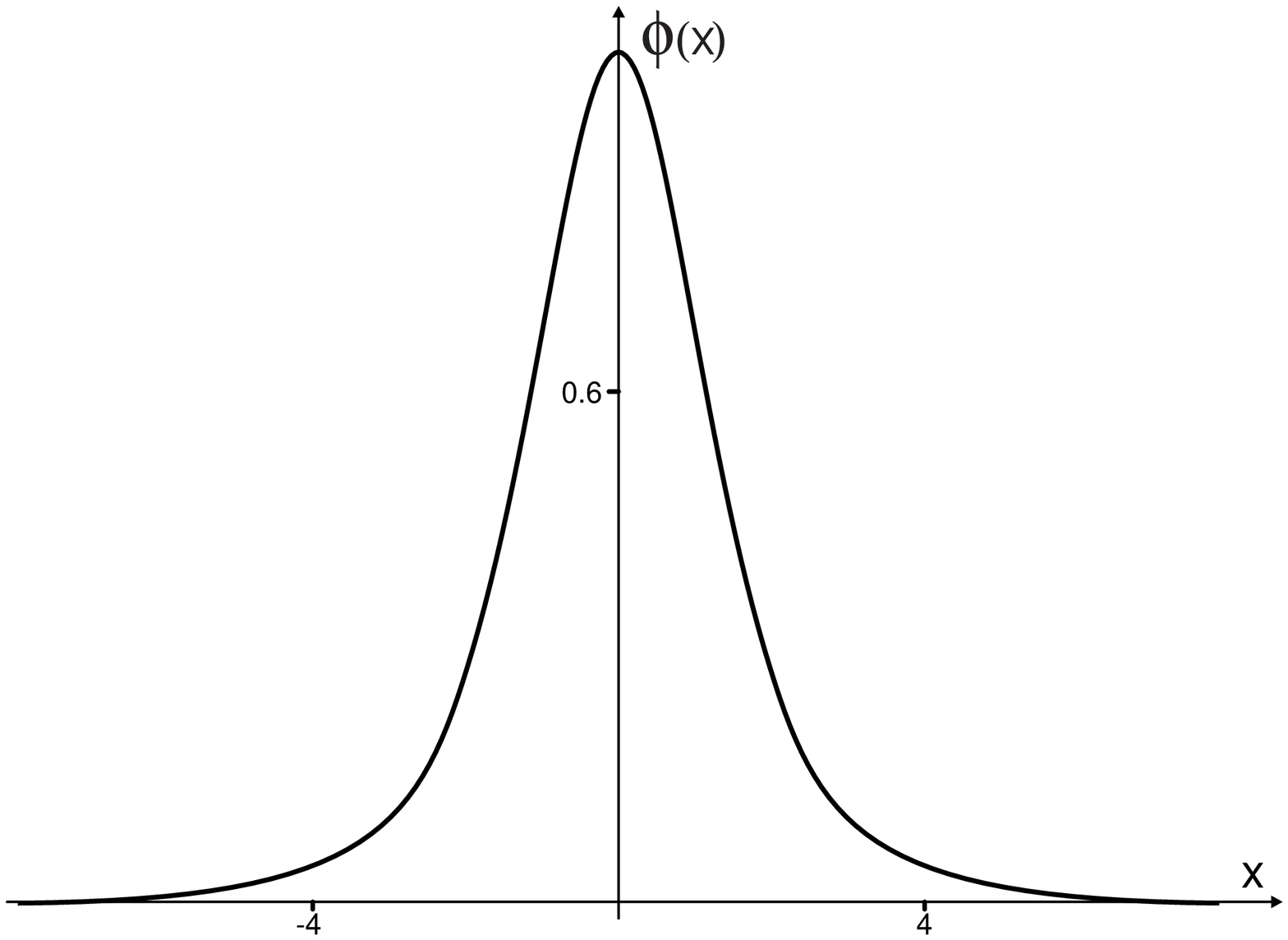}\vspace{.4cm}
\includegraphics[{width=5cm}]{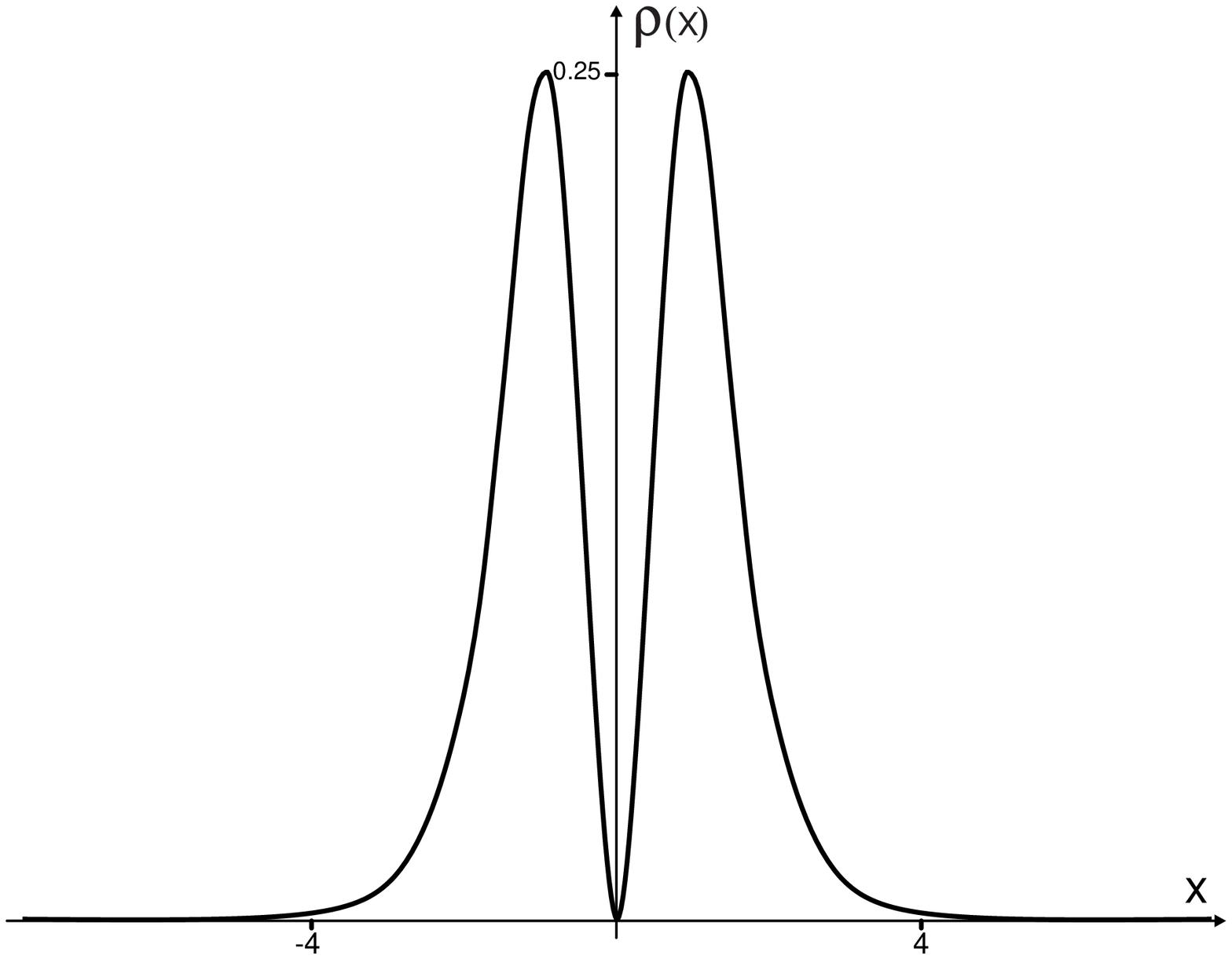}
\caption{Plots of the potential \eqref{phi3} (upper panel), the lump-like solution \eqref{phirho} (middle panel) and the energy density \eqref{phirho} (lower panel) for the $\phi^3$ model.}
\end{figure}

We can use the potential \eqref{phi3} to get 
\be\label{superpot1}
W_{\phi}=2\phi\sqrt{1-\phi}
\ee
and the square root of $1-\phi$ may induce error for $\phi\geq1,$ the region where the potential \eqref{phi3} is negative. However, this model has lump-like solution and energy density given by
\be
\phi(x)={\rm sech}^2(x);\;\;\;\rho(x)=3\,{\rm sech}^4(x)-3\,{\rm sech}^6(x)\label{phirho}
\ee
The amplitude and width of the non-topological bell-shaped solution are defined as: $A=|\phi(0)-\phi(\infty)|$ and $L=2|l|$, where $l$ is such that $\phi(l)=A/2.$ For the $\phi^3$ model we get $A_3=1$ and $L_3=2\,{\rm arcsech(\sqrt{2}/2)}.$ 

The lump-like solution \eqref{phirho} is positive and spans the interval $[0,1],$ never going to the region where the potential is negative. This means that the equations
\bes\ben
\frac{d\phi}{dx}&=&2\phi\sqrt{1-\phi};\;\;\;x<0
\\
\frac{d\phi}{dx}&=&-2\phi\sqrt{1-\phi};\;\;\;x>0
\een\ees
makes sense and may be used to respond for the lump-like solution. The other pair of equations describe the lump-like solution of the partner model, which is connected with the above one by changing $\phi\to-\phi.$ 

The model under investigation leads us to the expression 
\be
W(\phi)=\frac{4}{15}(2+3\phi)(1-\phi)^{3/2}
\ee
Since $W(0)=8/15$ and $W(1)=0,$ we get from \eqref{EL} that the energy of the lump-like solution is ${16}/{15}$. Note that this is the same value we get integrating the energy density using the explicit solution. To illustrate the model, in Fig.~1 we plot the potential, lump-like solution and the corresponding energy density.

We see that $W_{\phi\phi}=(2-3\phi)(1-\phi)^{-1/2}$. Thus, the potential \eqref{pq} of the Schroedinger-like Hamiltonian \eqref{ha} in this case is divergent at $x\to0$, making the factorization in terms of $S$ and $S^\dag$ ill-defined at the center of the lump-like structure.

Another example is the inverted $\phi^4$ model described by the potential
\be\label{p2}
V(\phi)=\frac12\phi^2(1-\phi^2)
\ee
This potential presents reflection symmetry, so it supports two distinct solutions, which are connected by the symmetry $\phi\to-\phi$. In this case we have
\be\label{superpot2}
W_{\phi}=\phi\sqrt{1-\phi^2}
\ee
and the first-order equations 
\be
\frac{d\phi}{dx}=-\phi\sqrt{1-\phi^2};\;\;x>0\;\;{\&}\;\; \frac{d\phi}{dx}=\phi\sqrt{1-\phi^2};\;x<0
\ee
and
\be
\frac{d\phi}{dx}=\phi\sqrt{1-\phi^2};\;x>0\;\;{\&}\;\; \frac{d\phi}{dx}=-\phi\sqrt{1-\phi^2};\;x<0
\ee
lead us to the solutions and energy density
\be
\phi(x)=\pm{\rm sech}(x);\;\;\;\rho(x)={\rm sech}^2(x)-{\rm sech}^4(x)\label{phirho4}
\ee
We use \eqref{superpot2} to get
\be\label{EL2}
W(\phi)=-\frac13(1-\phi^2)^{3/2}
\ee

\begin{figure}[ht!]
\includegraphics[{width=5cm}]{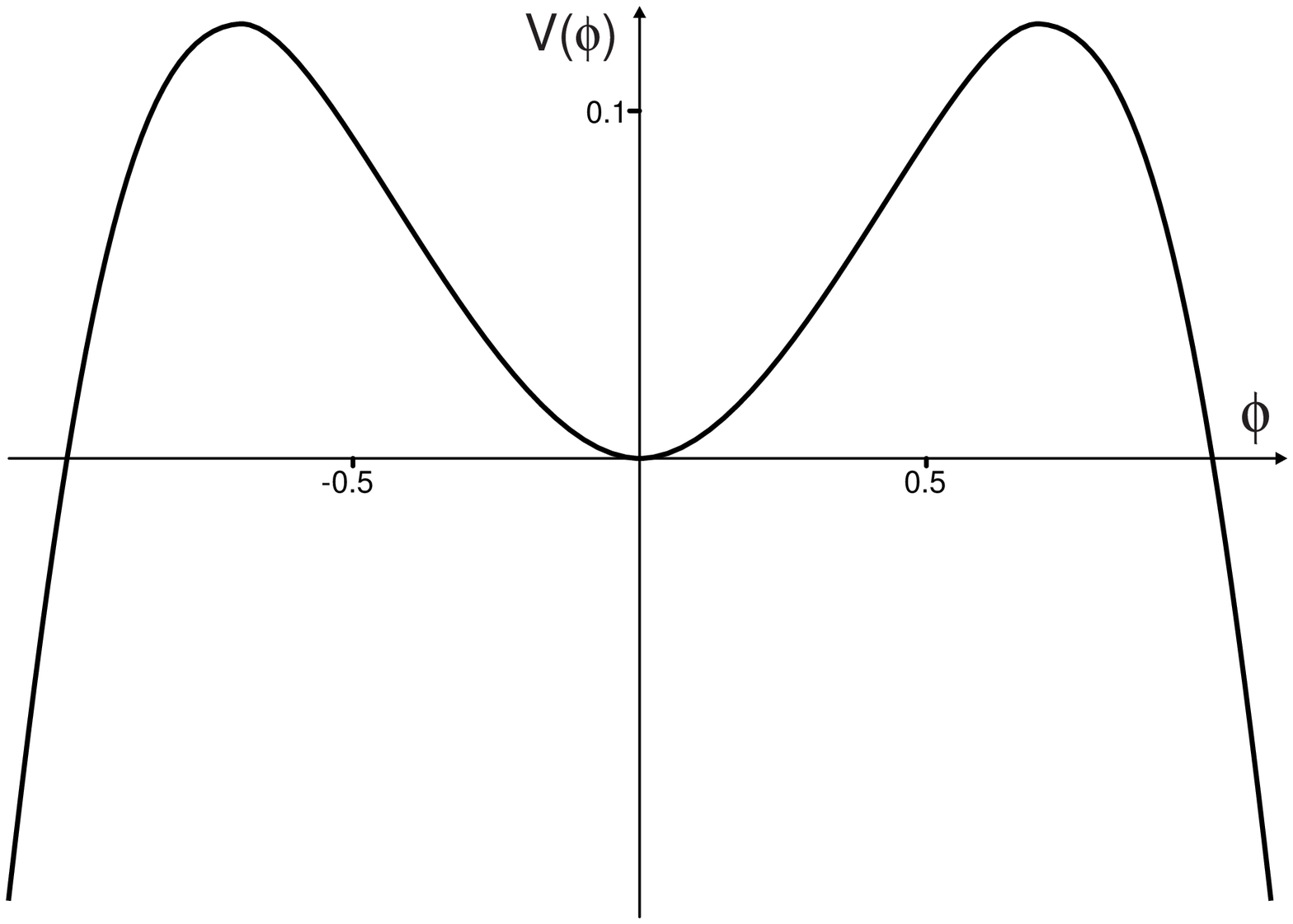}\vspace{.4cm}
\includegraphics[{width=5cm}]{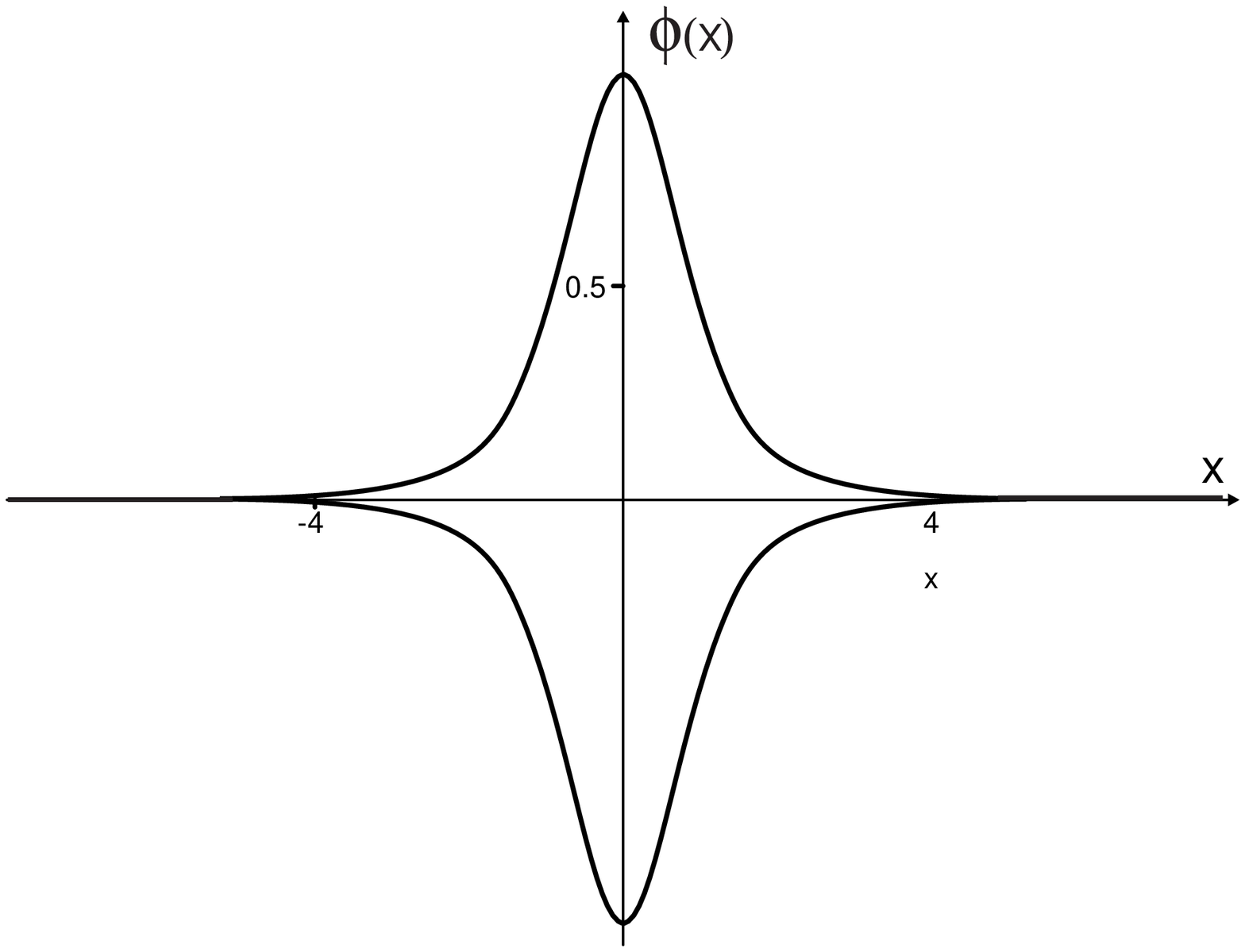}\vspace{.4cm}
\includegraphics[{width=5cm}]{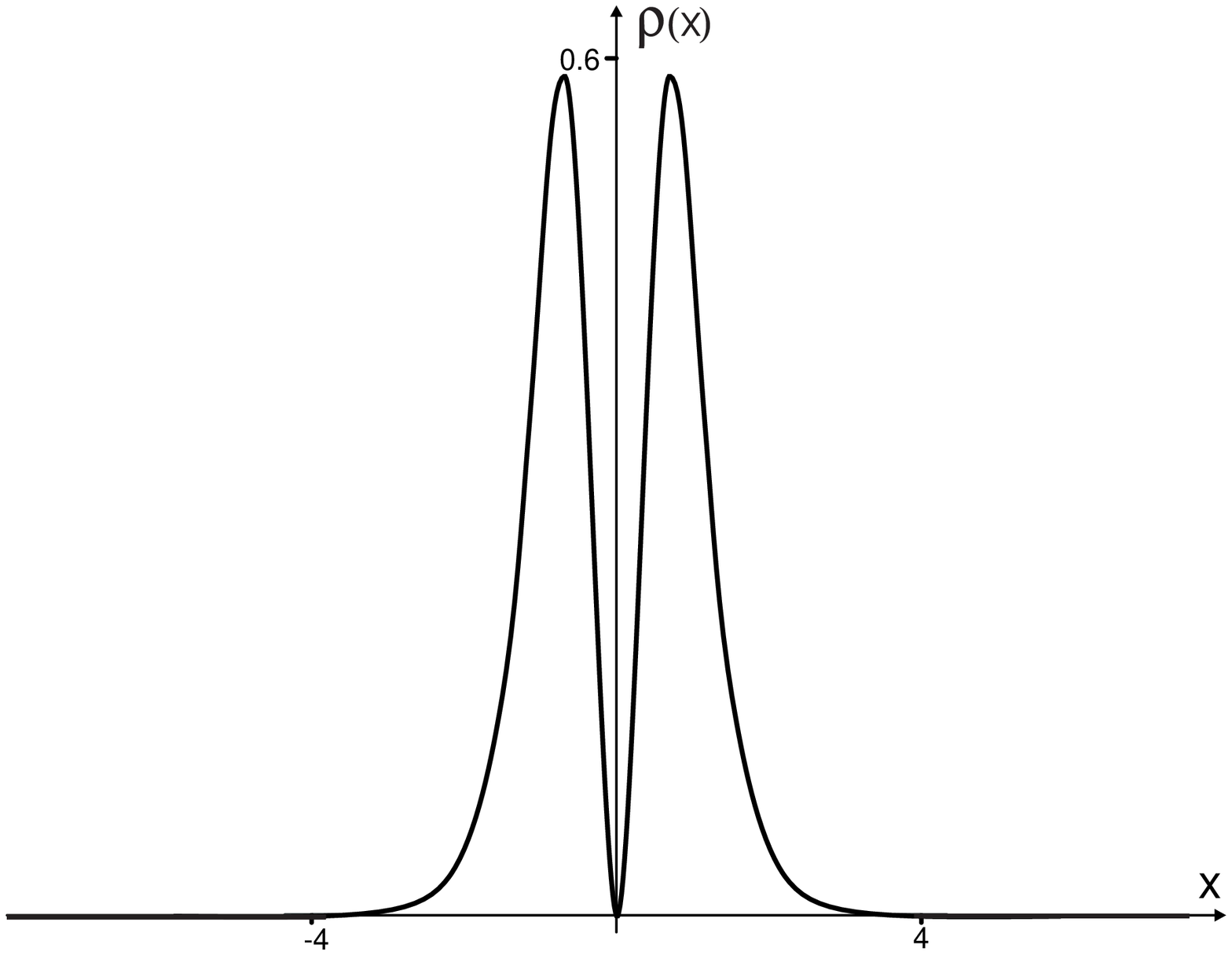}
\caption{Plots of the potential \eqref{p2} (upper panel), lump-like solutions \eqref{phirho4} (middle panel) and energy density \eqref{phirho4} (lower panel) for the inverted $\phi^4$ model.}
\end{figure}

We see that $W(0)=-1/3$ and $W(1)=0$, and so the energy of the lump-like solution is $E=2/3$. In Fig.~2 we plot the potential, lump-like solutions and the corresponding energy density. The maxima of the potential are located at $\phi^{\pm}_{max}=\pm\sqrt{2}/2$. The amplitude and width of the above lump-like solutions are $A_4=1$ and $L_4=2\,{\rm arcsech(1/2)}.$ For the inverted $\phi^4$ model we see that
\be
W_{\phi\phi}=\frac{2\phi^2-1}{(1-\phi^2)^{1/2}}
\ee
and the issue of instability is similar to the former case: the potential \eqref{pq} of the Schroedinger-like Hamiltonian \eqref{ha} is divergent at $x\to0$
in the present case, making the factorization in terms of $S$ and $S^\dag$ ill-defined at the center of the lump-like structure.

\section{Modified models}
\label{sec:spe}

We now focus attention on new models, which support a diversity of lump-like solutions. The models that we consider are modifications of standard models, and we study three distinct possibilities below. Here we notice that the new models are all described by a very small set of parameters. 

\subsection{The modified inverted $\phi^4$ model}

We consider the modified inverted $\phi^4$ model described by the potential
\be\label{potential1}
V=\frac12\phi^2(1+\phi)(a-\phi),
\ee
where $a$ is positive parameter. This new model is described by the single parameter $a,$ which nicely controls the profile of the non-topological solutions. For $a=1$ it reduces to the symmetric inverted $\phi^4$ model introduced in \cite{blm}. Here we can also use the reflection symmetry to construct another potential; this partner model can be solved in the same way we did for the $\phi^3$ and inverted $\phi^4$ models, so we will not comment on this anymore.

\begin{figure}[ht!]
\includegraphics[{width=6cm}]{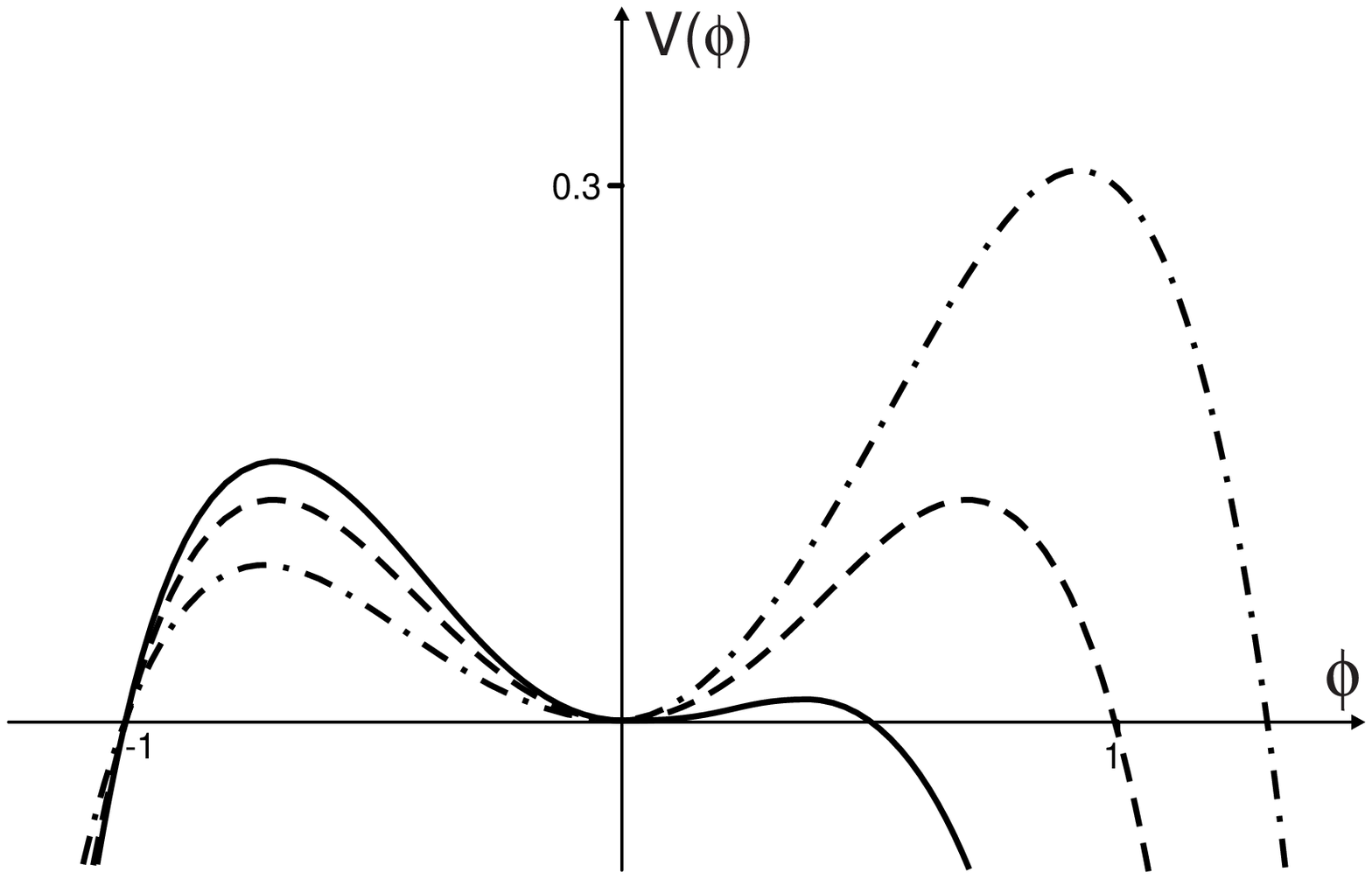}\vspace{.4cm}
\includegraphics[{width=6cm}]{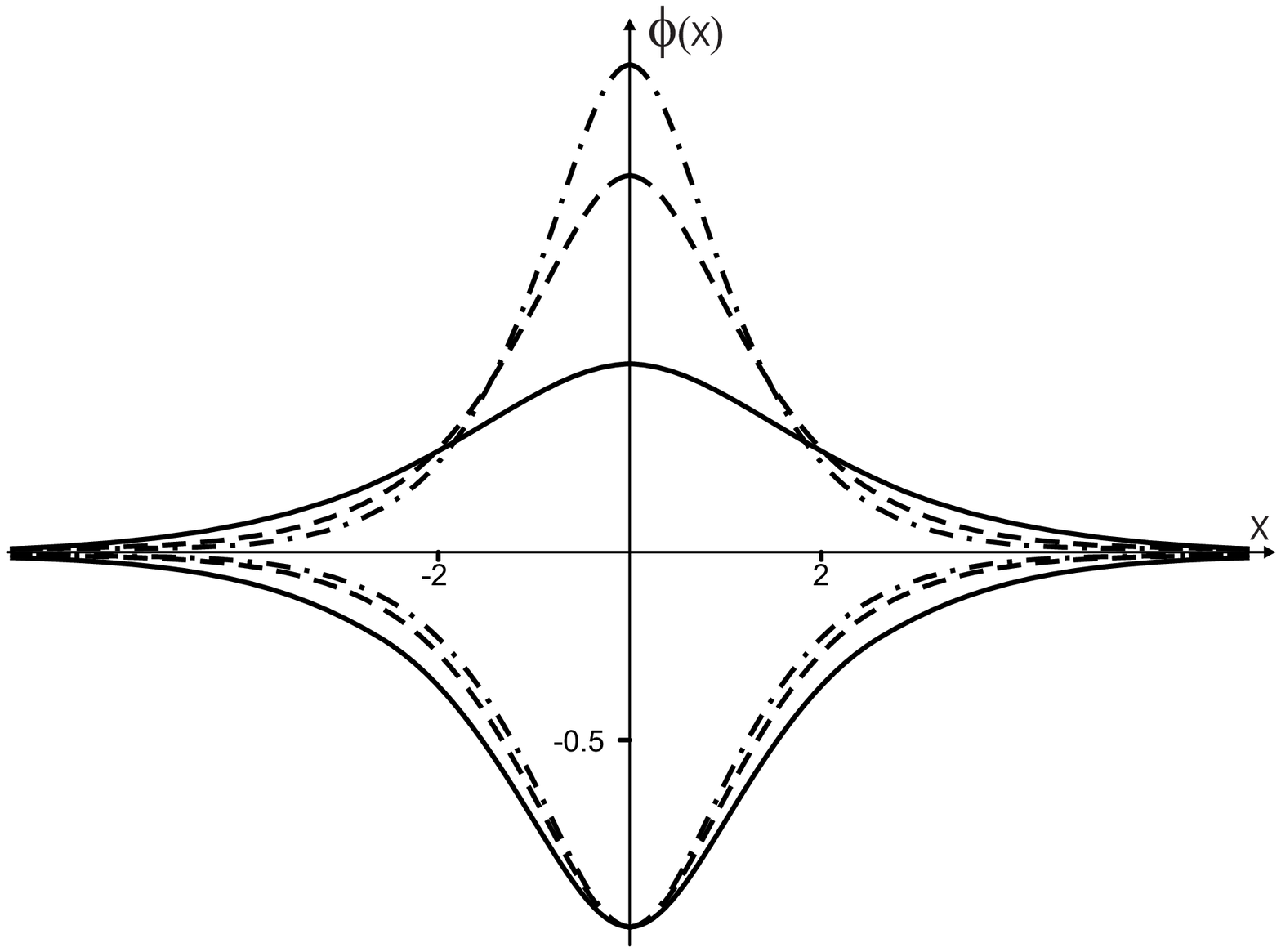}
\caption{Plots of the potential \eqref{potential1} (upper panel) and the lump solutions \eqref{lumps_1} (lower panel). The values of the parameter are $a=0.5$ (solid line), $a=1$ (dashed line), and $a=1.3$ (dash-dotted line).}
\end{figure}

The above potential has three zeros, one at $\bar\phi=0,$ which is the local minimum with $d^2V/d\phi^2=a$, and two others, at $\bar\phi_1=-1$ and at $\bar\phi_a=a$. In Fig.~3 we plot the potential for some values of the parameter. The maxima of the potential are located at
\be
\phi^{\pm}_{max}=-\frac38(a-1)\pm\frac18(9+14a+9a^2)^{\frac12}
\ee

We can use \eqref{EM11} and \eqref{EM12} to obtain the lump-like solutions
\bes\label{lumps_1}
\ben
\phi_1(x)&=&\frac{2a}{1-a-(1+a)\cosh(\sqrt{a}\,x)}
\\
\phi_a(x)&=&\frac{2a}{1-a+(1+a)\cosh(\sqrt{a}\,x)}\label{lumps_1a}
\een
\ees
The subscripts $1$ and $a$ are related with the two sectors of the potential. They reduce to the lumps $\phi(x)=\pm\,\,{\rm sech}(x)$, for $a=1$. The lump $\phi_1$ have the maximum $\phi_1(0)=-1$ and $\phi_a$ in $\phi_a(0)=a.$ Both lumps have asymptotic values $\phi_a(\pm \infty))=\phi_b(\pm \infty))=0.$ In Fig.~3, we show the lump-like solutions. These lumps have standard bell-shape profile: the $\phi_1$ solutions have constant amplitude and varying width, and the $\phi_a$ solutions have varying amplitude and width. The corresponding energy densities are given by
\bes\label{xxx}
\ben
\rho_1(x)&=&\frac{4a^3(1+a)^2\sinh^2(\sqrt{a}\,x)}{\left(1-a-(1+a)\cosh(\sqrt{a}\,x)\right)^4}
\\
\rho_a(x)&=&\frac{4a^3(1+a)^2\sinh^2(\sqrt{a}\,x)}{\left(1-a+(1+a)\cosh(\sqrt{a}\,x)\right)^4}
\een\ees
which are plotted in Fig.~4. The energies $E_1$ and $E_a$ corresponding to the above densities are given by
\bes\ben
E_1&=&\frac{\pi}{8}(1-a)(1+a)^{2}+E_a\label{E1}
\\
E_a&=&-\frac14(1-a)(1+a)^{2}\arctan\left(\!\sqrt{a}\right)\nonumber
\\
&&+\frac1{12}\sqrt{a}\left(3+2a+3{a}^{2}\right)\label{Ea}
\een\ees
We can check that the limit $a\to1$ sends both $E_1$ and $E_a$ to the same value $2/3,$ as it is shown in Fig.~4. The above lump-like solutions have amplitude and width given by $A^1_{4i}=1$ and $L^1_{4i}=(2/\sqrt{a})\,{\rm arccosh[(1-5a)/(1+a)]}$, and $A^a_{4i}=a$ and
$L^a_{4i}=(2/\sqrt{a})\,{\rm arccosh[(3+a)/(1+a)]}$. Here we notice that the limit $a\to1$ changes $A^a_{4i}\to A^1_{4i}$ and $L^a_{4i}\to L^1_{4i},$ as expected.

\begin{figure}[ht!]
\includegraphics[{width=7cm}]{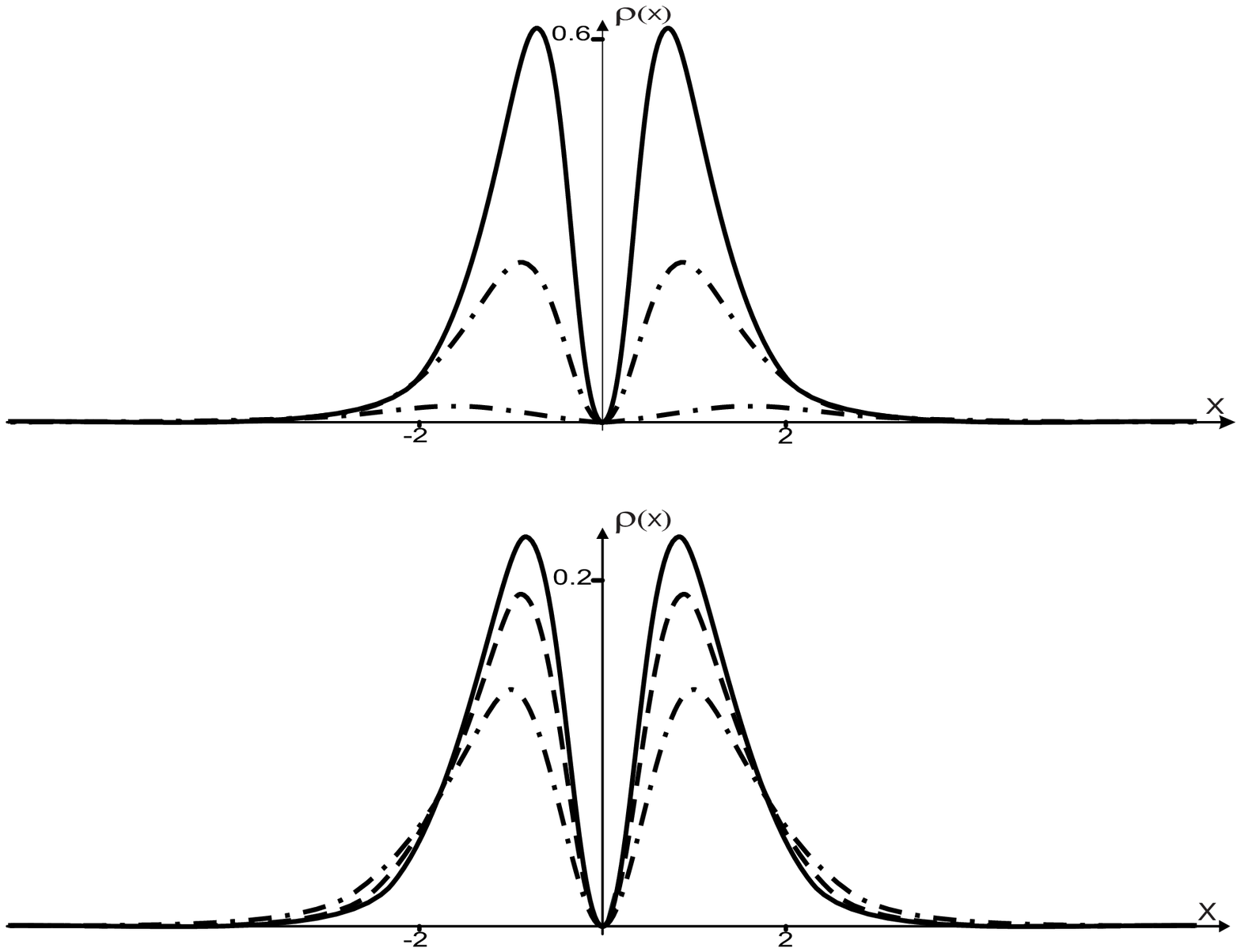}\vspace{.4cm}
\includegraphics[{width=6cm}]{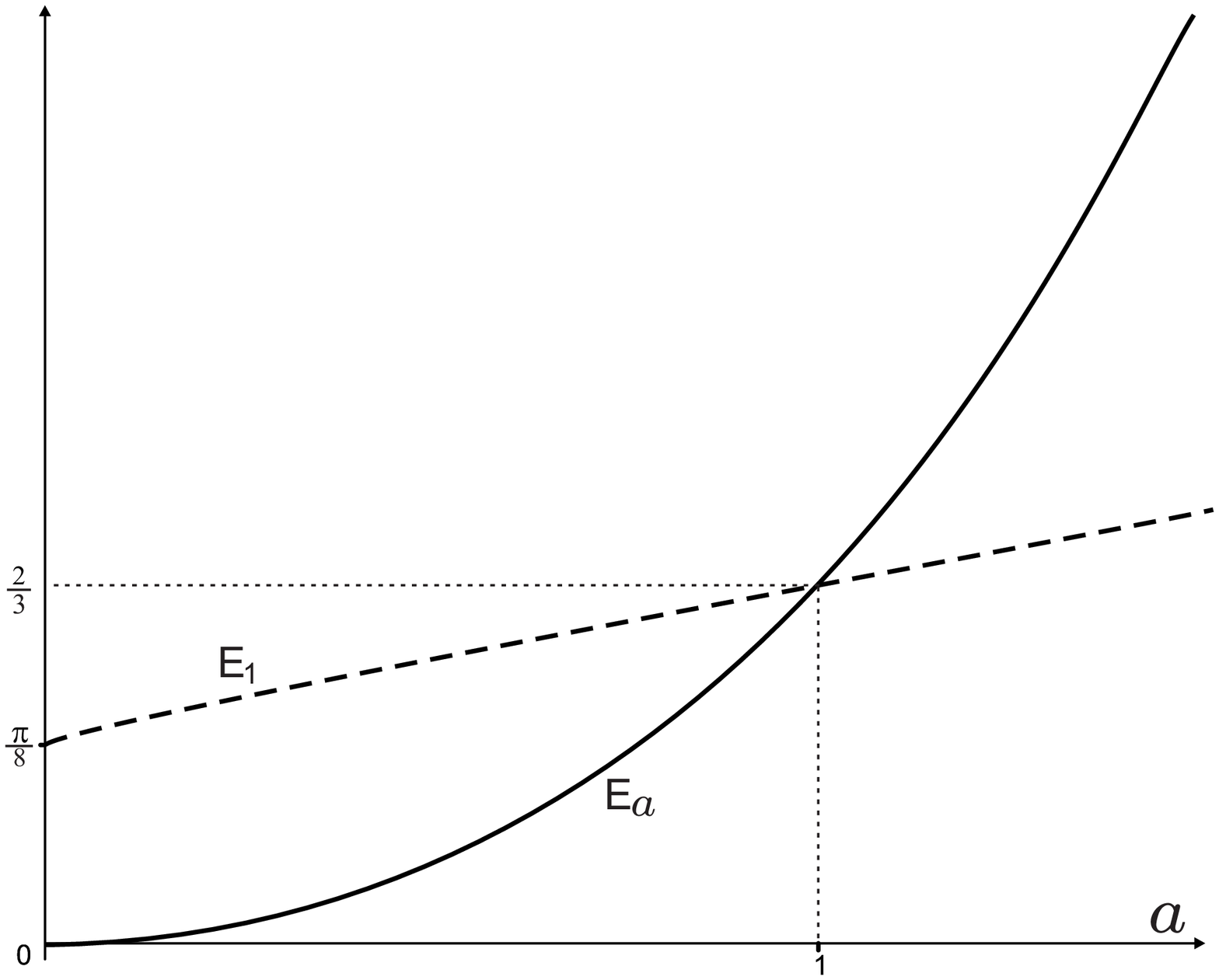}
\caption{Plots of the energy densities of the lump solutions $\phi_1$ (\ref{lumps_1}) (upper panel) and $\phi_a$ (\ref{lumps_1a}) (middle panel), for the values of the parameter $a=0.5$ (solid line), $a=1$ (dashed line), and $a=1.3$ (dash-dotted line). The lower panel shows the energies $E_1$ in \eqref{E1} (solid line) and $E_a$ in \eqref{Ea} (dashed line), as functions of the parameter $a$.}
\end{figure}

For this model defined by the potential (\ref{potential1}) we can write
\ben
W&\!\!=\!\!&\frac1{16}(a+1)^2(a-1){\rm arctan}\left(\frac{2\phi-a+1}{2G^{\frac12}(\phi)}\right)\nonumber
\\
&&+
\left(\frac{\phi^2}{3}-\frac{a-1}{12}\phi-\frac{(a-1)^2}{8}-\frac{a}{3}\right)G^{\frac12}(\phi)
\een
where $G(\phi)=a+(a-1)\phi-\phi^2$. It leads to the energies $E_1$ and $E_a$ given by \eqref{E1} and \eqref{Ea}, respectively.

The case with $a\to0$ is special. The minimum at $\phi=0$ becomes an inflection point and the potential changes to
\be\label{pota0}
V(\phi)=-\frac12\phi^3(1+\phi)
\ee
The lump-like solution has the form
\be
\phi(x)=-\frac{4}{x^2+4}
\ee
and the energy density is
\be
\rho(x)=\frac{64 x^2}{(x^2+4)^4}
\ee
The total energy of the solution is given by $E=\pi/8$. In Fig.~5 we plot the potential \eqref{pota0} to illustrate how it accommodates the inflection point. In Ref.~\cite{blm} we have constructed a model with kink-like solutions connecting two inflection points of the potential, and now we just constructed a model which supports lump-like solution with a single inflection point. As fas as we can see, this is the first time one construct such a solution.

\begin{figure}[ht!]
\includegraphics[{width=6cm}]{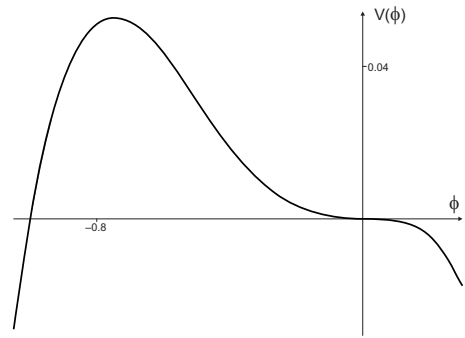}
\caption{Plot of the potential \eqref{potential1} in the limit $a\to0$, showing the presence of an inflection point at $\phi=0$.}
\end{figure}

\subsection{The modified $\phi^4$ model}

\begin{figure}[ht!]
\includegraphics[width=5cm]{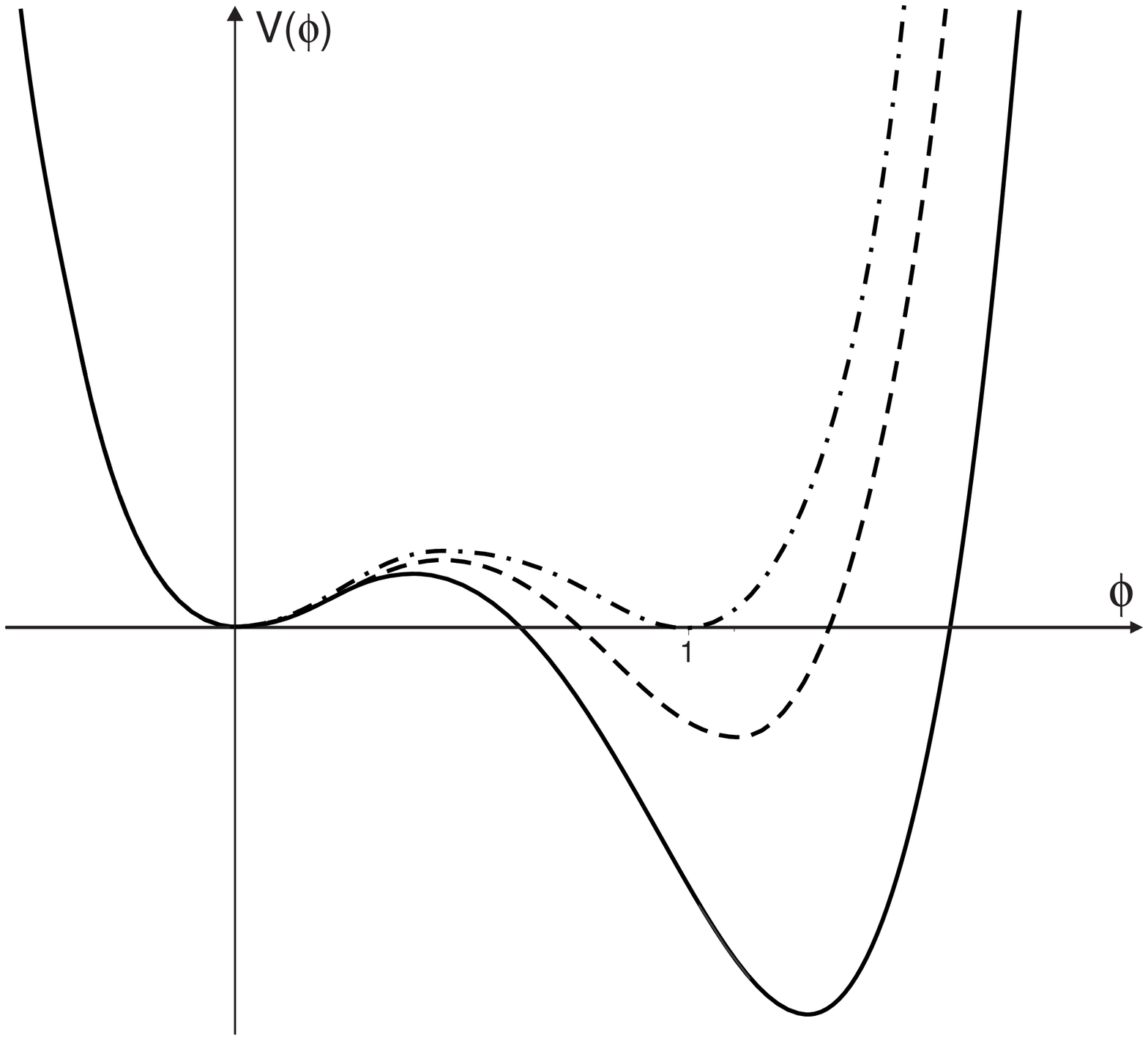}\vspace{.4cm}
\includegraphics[width=5cm]{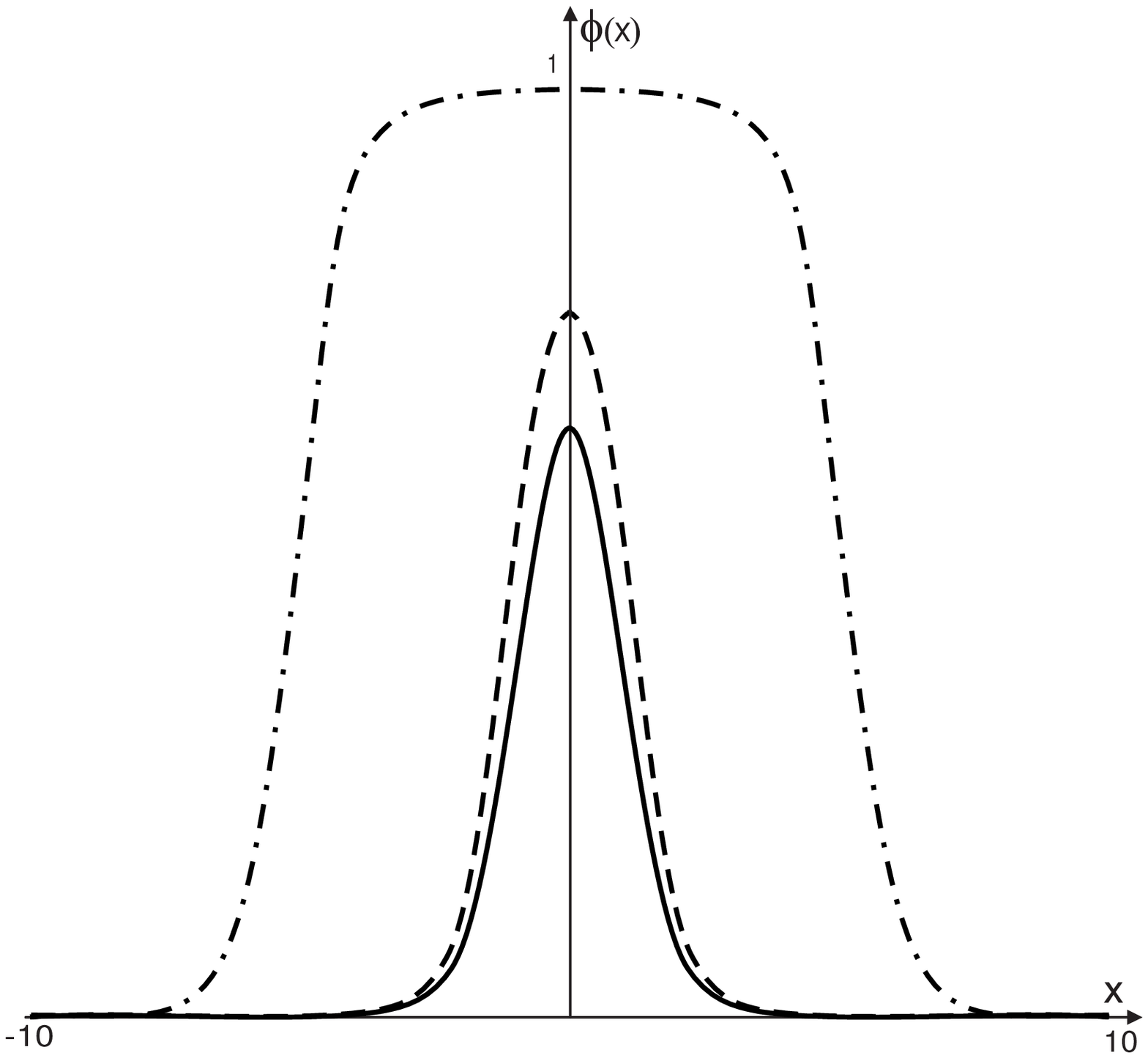}
\caption{Plots of the potential \eqref{pot2b} (upper panel) and the corresponding lump-like solution (lower panel) for $a=0.75$, $a=1$, and $a=4$, with solid, dashed, and dot-dashed lines, respectively.}\label{fb}
\end{figure}

Here we consider a model that presents large lump-like solutions described by a polynomial potential in $\phi$ up to fourth order, given by
\be\label{pot2a}
V(\phi)=2\phi^2(\phi-b)\left(\phi-\frac{b}{c}\right)
\ee
with the real parameters $b>0$ and $c\geq1$. In the case $c=1$ we obtain the $\phi^4$ model.
This potential has three zeros for $\phi=0$ (a local minimum), $\phi=b$ and $\phi=b/c$. There is a lump solution starting and ending at $\phi=0$, in the sector $0\leq\phi\leq b/c$, given by
\be
\phi(x)=\frac{b}{1+(1-c)\sinh^2\left(\sqrt{\frac{b^2}{c}}x\right)}
\ee
For $c\rightarrow1$ the lump width increases indefinitely. We can examine this feature in the simpler scenario, where we choose $b=\tanh(a)$ and $c=\tanh^2(a)$; in this case the potential \eqref{pot2a} takes the form
\be\label{pot2b}
V(\phi)=2\phi^2 \left(\phi - \tanh(a) \right) \left(\phi - \coth(a) \right)
\ee
It is now described by the single parameter $a>0,$ which nicely controls the profile of the non-topological solutions. We see that the limit $a\to\infty$ leads us back to the $\phi^4$ model: the redefinition $\phi=(\chi+1)/2$ gives 
\begin{equation}
V(\chi)=\frac18 \left(\chi^2-1 \right)^2
\end{equation}
which is the limit for $a\to\infty$ of the potential \eqref{pot2a}. We notice that in Fig.~{\ref{fb}} the value of the potential at the positive minimum is always negative, getting to zero in the limit $a\to\infty.$

\begin{figure}[ht!]
\includegraphics[width=5cm]{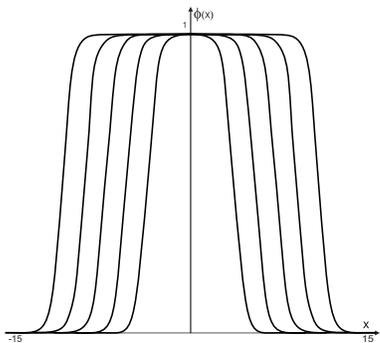}
\caption{Plots of the lump-like solution for $a=4$, $a=6$, $a=8$, $a=10$ and $a=12$, to illustrate that the width of the solution increases almost uniformly with increasing $a$.}
\end{figure}

The potential above has three critical points, one at $\phi_0=0$ and 
\bes
\begin{eqnarray}
\phi_{max}&=&\frac34 \coth(2a)-\frac14 \sqrt{9\coth^2(2a)-8}
\\
\phi_{min}&=&\frac34\coth(2a)+\frac14\sqrt{9\coth^2(2a)-8}	
\end{eqnarray}
\ees
The first point is a minimum, with $V(0)=0$ and $d^2V/d\phi^2(0)=2$. The other two points depend on $a,$ and for $a$ very large they become $\phi_{max}\approx (1/2)-3\exp(-4a)$ and $\phi_{min}\approx1+6\exp(-4a),$ going to $1/2$ and $1$ in the limit $a\to\infty$, respectively. The value of the potential at the maximum $\phi_{max}$ is always positive for $a\in[0,\infty),$ but at the minimum $\phi_{min}$ it is always negative, getting to zero asymptotically. This means that $V(\phi)$ has to cross the zero for $\phi$ somewhere in between $\phi_{max}$ and $\phi_{min},$ which we name $\phi_{back}=\tanh(a).$ Thus, we will ever find a non-topological lump-like solution, which starts at the minimum $\phi_0=0$ and returns to it after reaching $\phi_{back}=\tanh(a)$.

The lump-like solution centered at $x_0=0$ is given by
\begin{equation}\label{lump21}
\phi(x)=\frac12 \left(\tanh(x+a)-\tanh(x-a)\right),
\end{equation}
which can be seem as the addition of a kink at $x_0=-a$ and an anti-kink at $x_0=a.$ Although the addition of kink and anti-kink is sometimes used in the literature, we remind that here this possibility appears very naturally, as a lump-like solution of the model \eqref{pot2b}. The amplitude and width of the solution are $A_{4m}=\tanh(a)$ and
\be
L_{4m}=2\,{\rm arcsech}\left( \frac{1-\tanh(a)}{(2-\tanh^2(a))\tanh(a)}\right)^{\frac12}
\ee
For $a$ very large, we get $L_{4m}\approx 2a+\ln(2)+2e^{-2a},$ so it is a little larger then the distance between two kinks, one at $x=-a$ and the other at $x=a$. The maximum of the lump is $\phi_{back}=\tanh(a),$ which occurs at $x=0$, and so it approaches the unit as $a$ increases indefinitely. If we rewrite the lump-like solution in the form
\begin{equation}\label{lump22}
\phi(x)=\frac{\tanh(a)\,{\rm sech}^2(x)}{1-\tanh^2(a)\tanh^2(x)}
\end{equation}
we see that for $a$ very small it becomes $\phi=a \,{\rm sech}(x)^2$, which is similar to the solution found in the $\phi^3$ model, so it solves the model
\begin{equation}
V(\phi)=2\phi^2\left(1-\frac{\phi}{a}\right) \label{Vphi3}
\end{equation}

As a bonus of the above description, we see that the width of the lump increases for increasing $a,$ depicting a flat plateau around its maximum. The usual bell-shape profile of the lump-like solution then changes to a different profile, with the bell shape developing a plateau at its top. This form is unusual and we believe it could perhaps be used in applications where the bell-shape solution finds some utility. This fact will be explored elsewhere.

We see that the width of the plateau at the top of the hill increases with $a$ increasing. The energy density gets two maxima, at the values $-a$ e $+a,$ and for $a$ very large the two hills disconnect. The analytic expression for the energy density is
\begin{equation}\label{rho3m}
\rho(x)=\frac14 \left[\tanh^2(x+a)-\tanh^2(x-a)\right]^2
\end{equation}
and the energy is 
\ben
E&=&2\,\int^\infty_{-\infty} dy \,\,\frac{1}{4} {\rm sech}	(y)^4 \nonumber
\\
&&-2\,\int^\infty_{-\infty} dy \,\,\frac{1}{4} {\rm sech}	(y-a)^2 {\rm sech}(y+a)^2 
\een
In the above expression for the energy, the first integral does not depend on $a;$ it gives the sum of the energies corresponding to each of of the two kinks.
The second term depends on $a,$ and it is always lower then the first integral, reaching the same value for $a=0$. We calculate the integrals to get
\begin{equation}\label{ener2}
E=\frac23-2\, {\rm cossech}^2(2a)\bigl( 2a\,{\rm coth}(2a)-1\bigr)
\end{equation}
It is zero for $a=0$, and it goes to $2/3$ for $a\to\infty.$ In this case $W(\phi)$ is given by
\ben
W&\!\!=\!\!&\left(\frac23-\frac{A^2}{4}-\frac{A}{6}\phi+\frac23\phi^2\right)G^{\frac12}(\phi)\nonumber
\\
&&\!\!+A\!\left(\!1-\frac{A^2}{4}\!\right)\!{\rm arctanh}\!\left(\!\frac{G^{\frac12}(\phi)-1-\frac{A}{2}+\phi}{G^{\frac12}(\phi)+1-\frac{A}{2}+\phi}\!\right)
\een
where $G(\phi)=1-A\phi+\phi^2,$ and $A=\tanh(a)+{\rm coth}(a).$ It can be used to obtain the above value of the energy.

\begin{figure}[ht!]
\includegraphics[width=7cm]{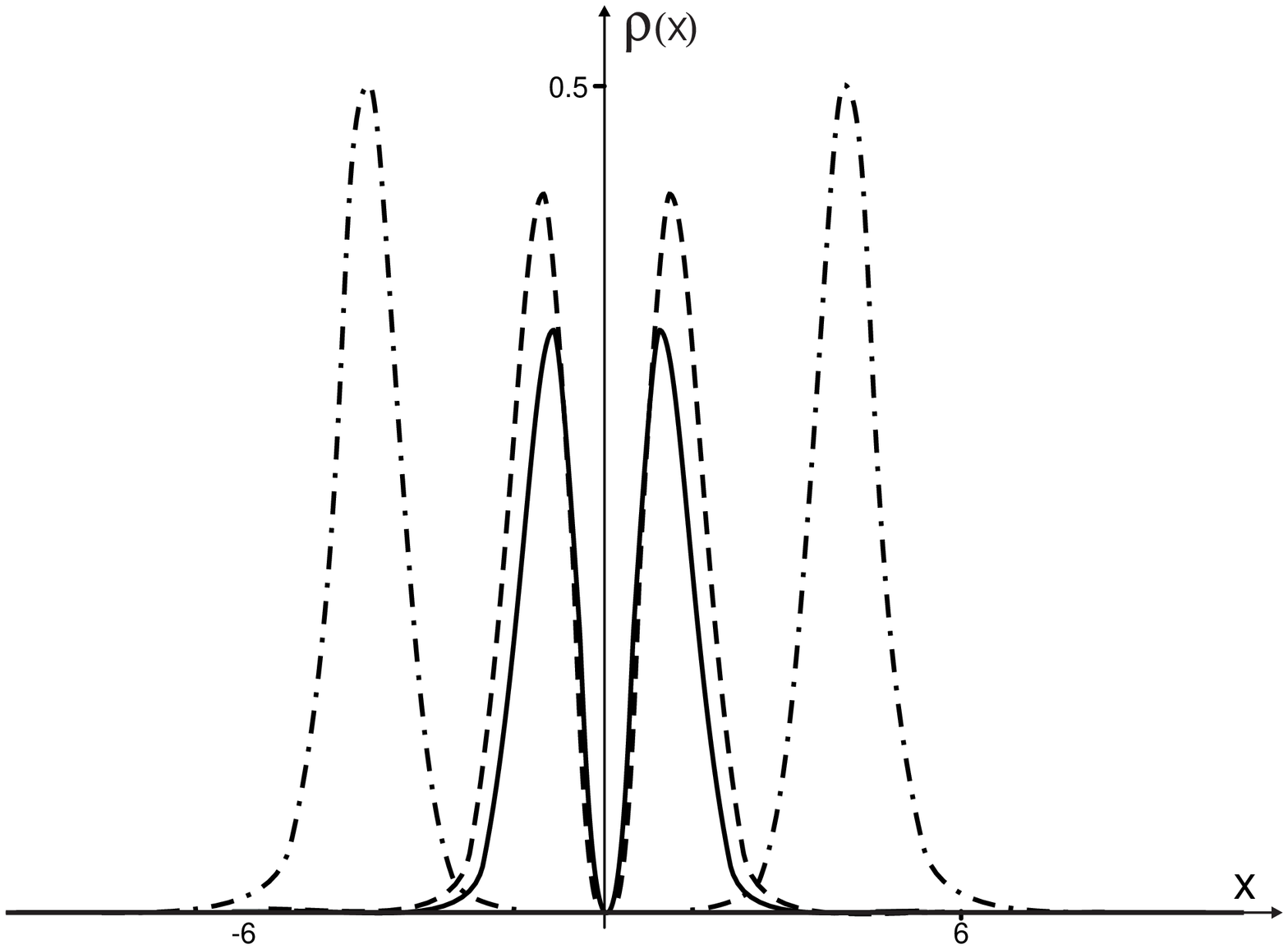}\vspace{.4cm}
\includegraphics[width=5cm]{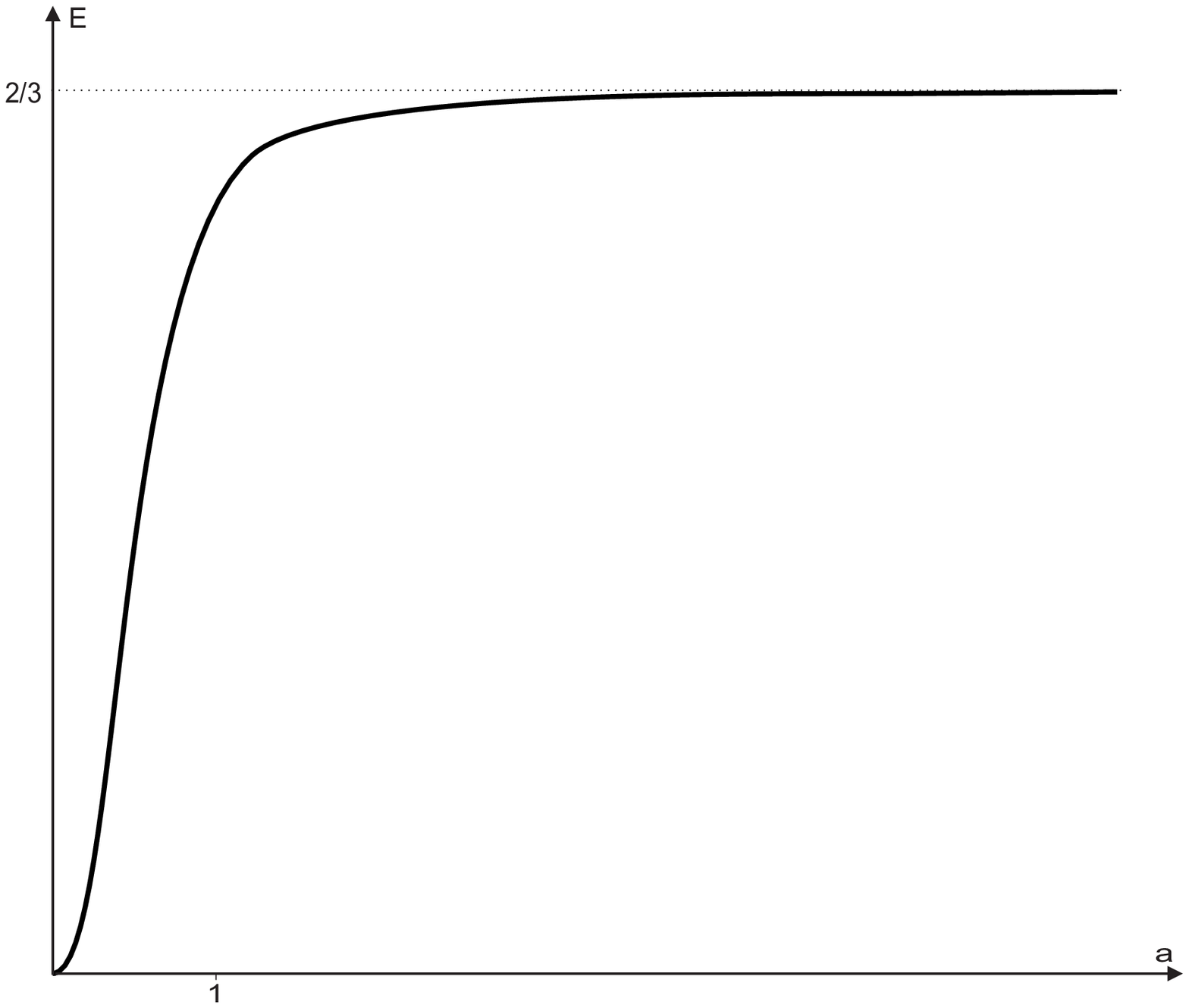}
\caption{Plots of the energy density \eqref{rho3m} (upper panel) for $a=0.75$ (solid line), $a=1$ (dashed line) and $a=4$ (dot-dashed line), and the energy \eqref{ener2} (lower panel) as a function of $a.$}
\end{figure}

\subsection{The modified $\phi^3$ model}

\begin{figure}[ht!]
\includegraphics[{width=5cm}]{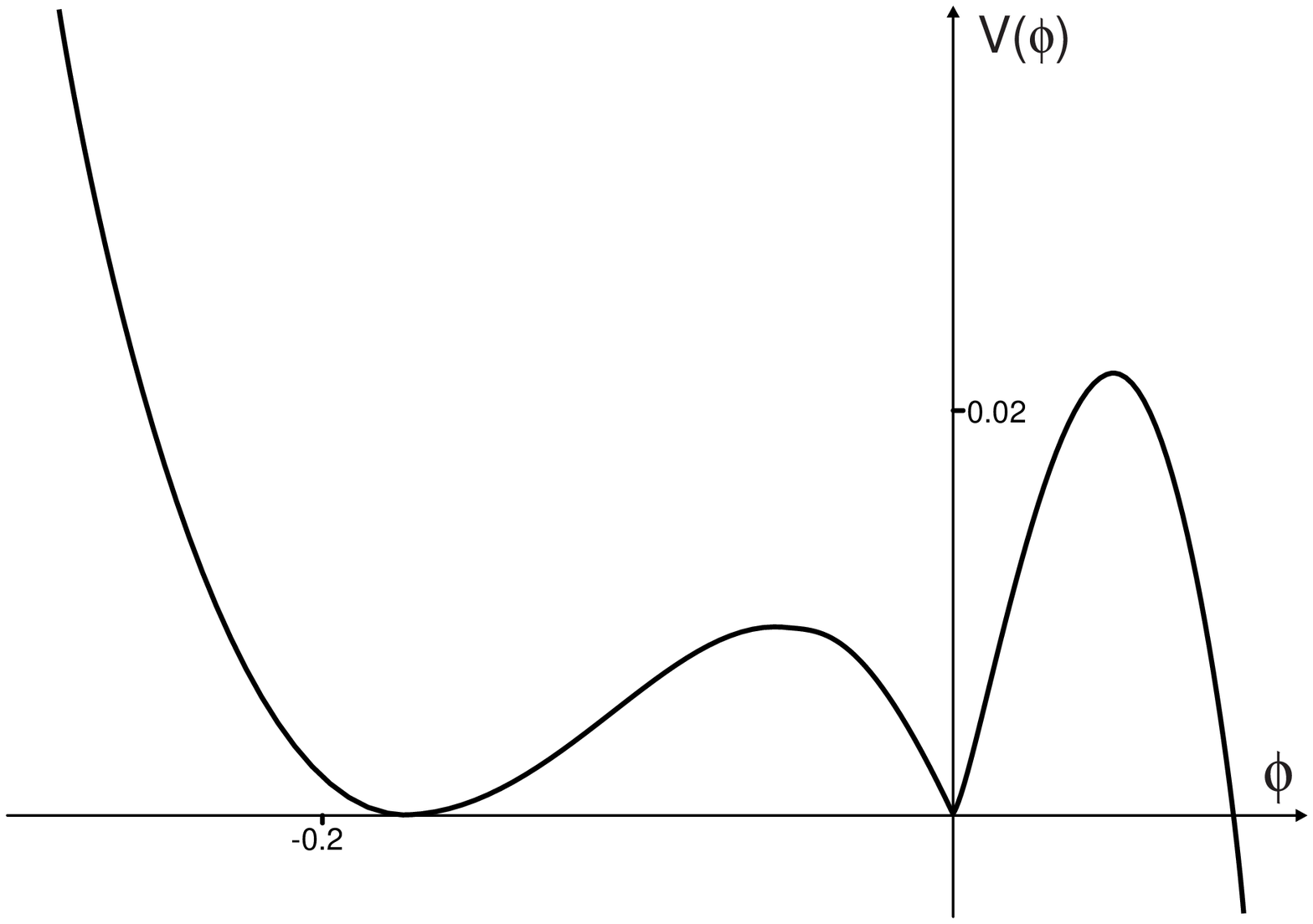}\vspace{.4cm}
\includegraphics[{width=5cm}]{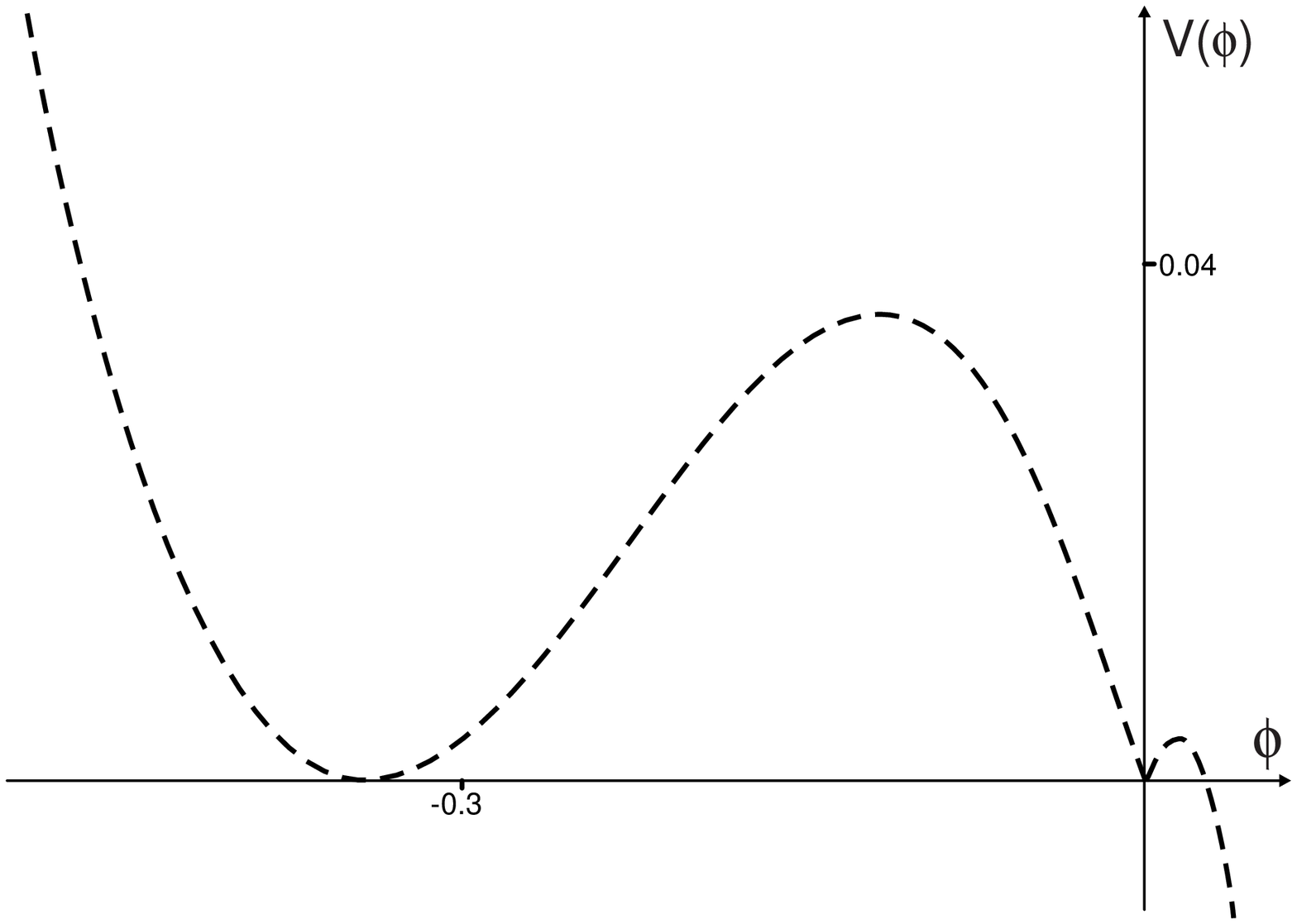}\vspace{.2cm}
\includegraphics[{width=5cm}]{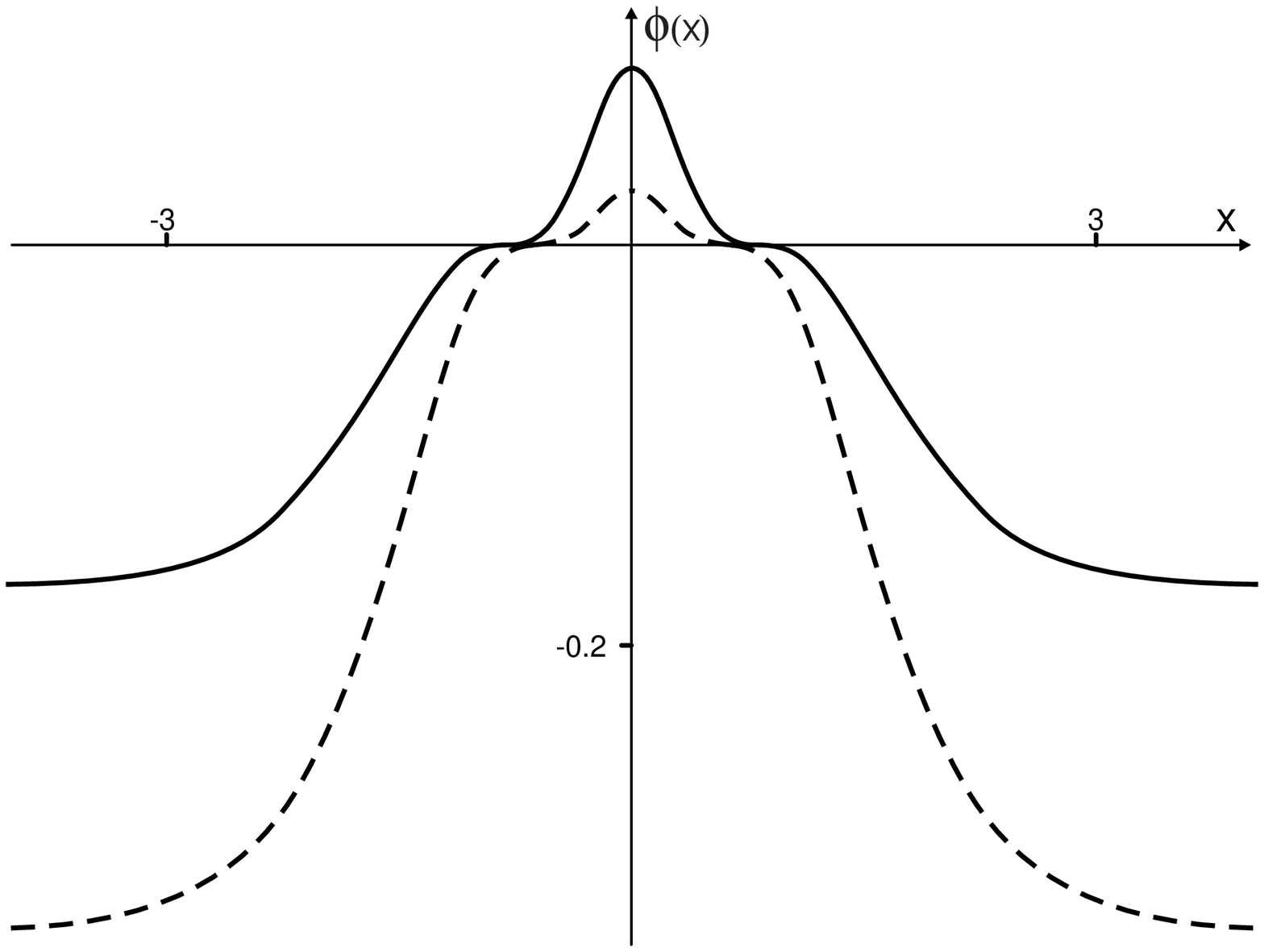}\vspace{.4cm}
\includegraphics[{width=5cm}]{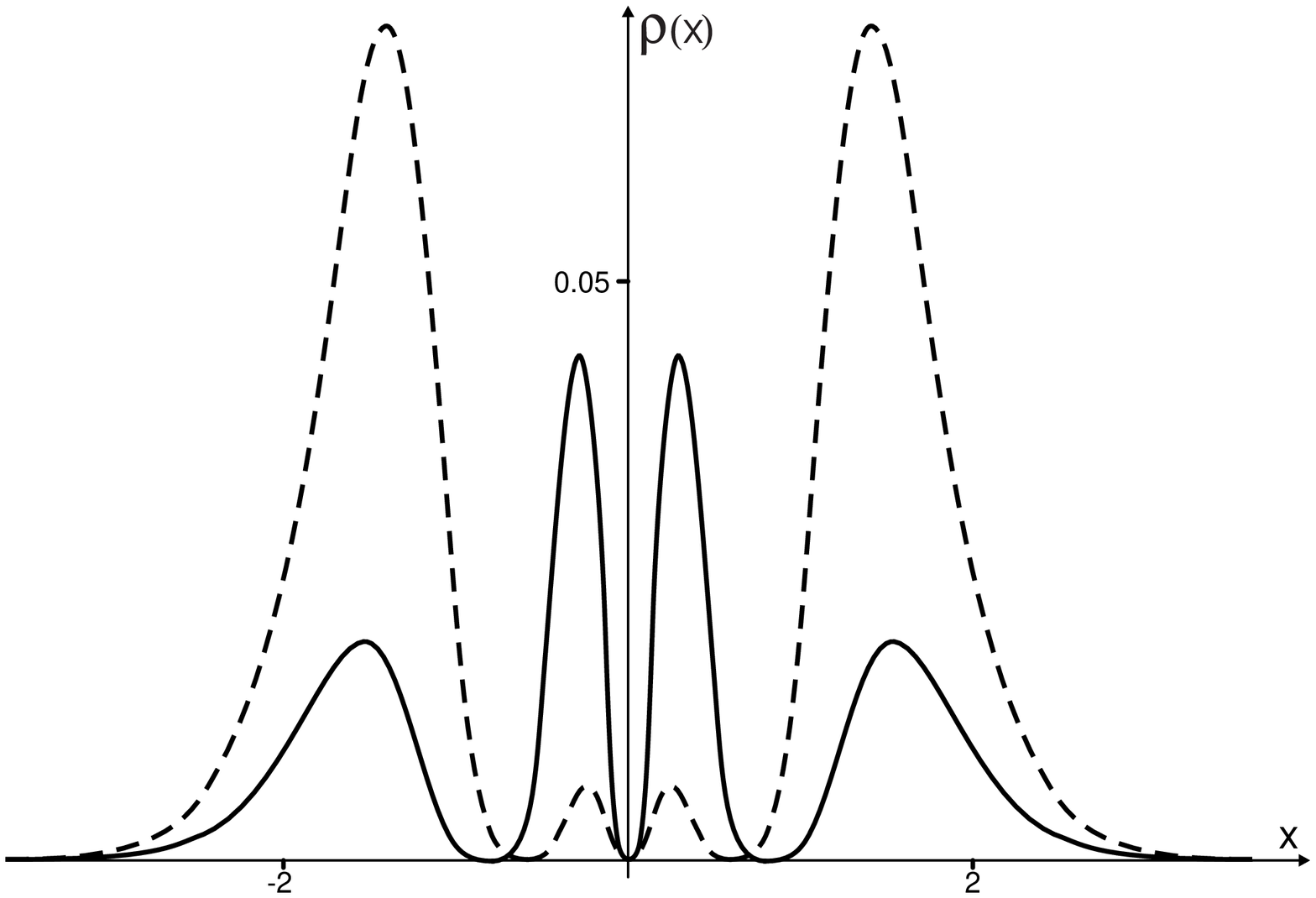}
\caption{Plots of the potential \eqref{dlphi3} for $p=3$ and $a=0.554$ (upper panel) and $a=0.70$ (middle panel, upper side), and the related 2-lump solution (middle panel, lower side) and the corresponding energy density (lower panel).}\label{dlphi3a}
\end{figure}

We also consider the model described by the potential
\be\label{dlphi3}
V(\phi)=2p^2\phi^{2-\frac2p}\left(1-a-\phi^{\frac1p}\right)\left(a+\phi^{\frac1p}\right)^2
\ee
where $a$ and $p$ are positive parameters, with $a\in[0,1]$ and $p=1,3,5,...$, is an odd integer. This new model is described by the two parameters $a$ and $p,$ which control interesting new features of the non-topological solutions. For $a=0$ and $p=1$ the model reduces to the $\phi^3$ model described before in \eqref{phi3}. This potential has three zeros, two of then being local minima, at $\bar\phi_0=0$ and $\bar\phi_1=-a^p,$ and one at $\bar\phi_2=(1-a)^p.$ Here we have the lump-like solutions
\be\label{S3}
\phi(x)=\left({\rm sech}^2(x)-a\right)^p
\ee
and the corresponding energy densities
\be
\rho(x)=4p^2{\rm sech}^4(x)\,{\tanh^2(x)}\bigl({\rm sech}^2(x)-a\bigr)^{2p-2}
\ee
In Fig.~(\ref{dlphi3a}) we plot the potentials and the related solutions and energy densities for $p=3,$ and $a=0.554$ and $0.70$. The maxima of the potential are located at
\be
\phi^{\pm}_{max}=\pm\left(\frac{A^{\frac12}\pm 4ap-a-2p}{2(2p+1)}\right)^p
\ee
where $A=4p^2+9a^2-4ap-4a.$

The corresponding lump-like solutions are special. We see that their energy densities develop four maxima, instead of the two maxima that usually appear in the standard case. In fact, they are similar to the 2-kink solutions found before in \cite{bmm}, so we name then 2-lump. The profile of the 2-lump solutions is unusual; it seems to be a lump on top of another lump. It has a maximum at $x=0,$ goes to zero at $\pm(1-a)^p,$ and gets to the value $-a^p$ at the limits $x\to\pm\infty.$

The amplitude and width of these solutions are given by $A_{3m}=a^p+(1-a)^p$ and $L_{3m}={\rm arcsech}\sqrt{\alpha}$, where
\be
\alpha=a+a\left(\frac12\left(1+\left(1-\frac1a\right)^p\right)\right)^\frac1p
\ee
Since the 2-lump solution is a kind of composite solution, we can also introduce amplitude and width for the top and bottom lumps separately: they are given by
$A^t_{3m}=(1-a)^p$ and $L^t_{3m}={\rm arcsech}\sqrt{\alpha_t}$, and $A^b_{3m}=a^p$ and $L^b_{3m}={\rm arcsech}\sqrt{\alpha_b}$, with
$\alpha_t=2^{-1/p}+a(1-2^{-1/p})$ and $\alpha_b=a(1+2^{-1/p})$.

We do not show $W(\phi)$ in general since it is an awkward expression. But we can make $p=3$ to obtain
\be
W_3\!=\!-\frac92 a\!\left(\!a^2\!+\!\frac12\right)\!{\rm arcsin}\left(\!a\!-\!\phi^{\frac13}\right)\!+\!F_3(\phi)G_3^{\frac12}(\phi)
\ee
where
\ben
F_3(\phi)&=&A+B\phi^{\frac13}+C\phi^{\frac23}+\frac{3}{35}a\left(a^2-\frac{11}{2}\right)\phi\nonumber
\\
&&+\frac{3}{35}(a^2-3)\phi^{\frac43}-\frac{12}{7}a\phi^{\frac53}+\frac{9}{7}\phi^2
\een
and
\bes\ben
A&=&\frac{3}{35}a^6-\frac{6}{7}a^4-\frac{741}{140}a^2-\frac{24}{35}
\\
B&=&\frac{3}{35}a^5-\frac{27}{35}a^3-\frac{219}{140}a
\\
C&=&\frac{3}{35}a^4-\frac{9}{14}a^2-\frac{12}{35}
\een\ees
and
\be
G_3(\phi)=1-a^2+2\phi^{\frac13}-\phi^{\frac23}
\ee
In this case the energy becomes
\be
E_3(a)=\frac{2048}{1001}-\frac{4096}{385}a+\frac{768}{35}a^2-\frac{768}{35}a^3+\frac{48}{5}a^4
\ee
It is of the fourth degree in $a,$ and we can show that in general the energy for $p$ arbitrary is of the $2(p-1)$ degree in $a.$   
For $p=3,$ the minimum value of the energy is $0.060,$ and it occurs at $a=0.554.$ For $p=5,$ the minimum is $0.004$ and occurs at $a=0.538.$

The case $a=1$ is interesting. The potential changes to
\be
V(\phi)=-2p^2\phi^{2}\left(\phi^{-\frac{1}{2p}}+\phi^{\frac{1}{2p}}\right)^2
\ee
and the lump-like solution \eqref{S3} becomes
\be
\phi(x)=-\tanh^{2p}(x)
\ee 
It has amplitude and width given by $A^1_{3m}=1/2$ and $L^1_{3m}=2\,{\rm arctanh(1/2)^{1/2p}}.$ It is a special case which was investigated before in \cite{bmm}. Here the energy is given by
\be
E_p(1)=\frac{16p^2}{16p^2-1}
\ee
This solution has a plateau which increases with $p,$ and in the limit $p\to\infty$ the energy goes to $1$.

\section{Ending comments}

In this work we have studied several models which support lump-like solutions of distinct profiles, not present in the related literature. We started the investigations with the general formalism, and there we have built a first-order framework to study non-topological or lump-like defect structures.
This extension is interesting, since it leads to first-order differential equations which are somehow easier to solve. As a bonus, it also furnishes a simple and direct way to obtain the energy of the corresponding solution.

The first-order framework for non topological or lump-like solutions has inspired us to investigate other more complex models. We have introduced three distinct classes of models, from which we could obtain several lump-like solutions. Some of these new solutions are characterized by having varying amplitude and width, others have appeared with the usual bell-shape form changed to engender a flat plateau on its top or to give rise to a bell-shape solution on top of another bell-shape structure. The present investigations add several results to the somehow hard subject of proposing new models and finding explicit analytical solutions of non-topological profile. The next steps can follow two distinct routes: one concerning further generalizations of the above scenario, changing the discrete symmetry to a continuum symmetry, making it global or local, Abelian or non-Abelian, with the inclusion of complex scalars and gauge fields; another route relies on the use of the above results to study applications of specific interest in diverse areas of non-linear science. These and other issues are presently under consideration, and we hope to report on them in the near future.

We would like to thank CAPES, CNPq, and MCT-CNPq-FAPESQ for partial financial support.


\end{document}